\documentclass[aps,prx,twocolumn,showpacs,superscriptaddress,floatfix,footinbib]{revtex4-2}

\usepackage{graphicx,graphics}
\usepackage{dcolumn}
\usepackage{amsmath,amssymb,amsfonts}
\usepackage{latexsym,verbatim}
\usepackage{bm}
\usepackage{lipsum} 
\usepackage{dsfont}
\usepackage{cancel}
\usepackage{breqn}
\usepackage{bbm} 
\usepackage{algorithm}
\usepackage[noend]{algpseudocode}

\makeatletter
\let\cat@comma@active\@empty
\makeatother

\usepackage[toc,page]{appendix}
\usepackage{color}
\usepackage{ulem}
\usepackage{tikz}
\usepackage[breaklinks=true,colorlinks,citecolor=blue,linkcolor=blue,urlcolor=blue]{hyperref}
\usepackage{tabularx}
\usepackage{array, multirow, bigdelim, makecell, booktabs} 
\graphicspath{fig}

\newcommand{\ourAnsatz}{l-USC}
\newcommand{\denseAnsatz}{d-USC}

\newenvironment{diagram}
{
\begin{tikzpicture}[baseline = (X.base),every node/.style={scale=0.7},scale=.55]
}
{
\end{tikzpicture}
}

\begin{document}

\title{Time Evolution of Uniform Sequential Circuits}
\author{Nikita~Astrakhantsev}
\email[]{nikita.astrakhantsev@physik.uzh.ch}
\affiliation{Department of Physics, University of Zurich, Winterthurerstrasse 190, CH-8057 Z\"{u}rich, Switzerland}

\author{Sheng-Hsuan Lin}
\affiliation{Technical University of Munich, TUM School of Natural Sciences, Physics Department, 85748 Garching, Germany}

\author{Frank Pollmann}
\affiliation{Technical University of Munich, TUM School of Natural Sciences, Physics Department, 85748 Garching, Germany}
\affiliation{Munich Center for Quantum Science and Technology (MCQST), Schellingstr. 4, 80799 M{\"u}nchen, Germany}

\author{Adam Smith}
\affiliation{School of Physics and Astronomy, University of Nottingham, Nottingham, NG7 2RD, UK}
\affiliation{Centre for the Mathematics and Theoretical Physics of Quantum Non-Equilibrium Systems, University of Nottingham, Nottingham, NG7 2RD, UK}

\begin{abstract}
Simulating time evolution of generic quantum many-body systems using classical numerical approaches has an exponentially growing cost either with evolution time or with the system size. In this work, we present a polynomially scaling hybrid quantum-classical algorithm for time evolving a one-dimensional uniform system in the thermodynamic limit. This algorithm uses a layered uniform sequential quantum circuit as a variational ansatz to represent infinite translation-invariant quantum states. We show numerically that this ansatz requires a number of parameters polynomial in the simulation time for a given accuracy. Furthermore, this favourable scaling of the ansatz is maintained during our variational evolution algorithm. All steps of the hybrid optimization are designed with near-term digital quantum computers in mind. After benchmarking the evolution algorithm on a classical computer, we demonstrate the measurement of observables of this uniform state using a finite number of qubits on a cloud-based quantum processing unit. With more efficient tensor contraction schemes, this algorithm may also offer improvements as a classical numerical algorithm.
\end{abstract}
\maketitle

\section{Introduction}

Performing time evolution of quantum states far-from-equilibrium represents a challenging problem for the contemporary study of quantum matter.
Beyond rare analytically tractable settings~\cite{PhysRevLett.110.257203,de2015relaxation,piroli2019integrable}, exact numerical methods scale with the Hilbert space, whose dimension scales exponentially with the number { of} degrees of freedom.
Approximate methods based on tensor networks have released this constraint with the cost of exponential scaling with bipartite entanglement entropy~\cite{PhysRevLett.69.2863,Schollw_ck_2011}.
While this offers dramatic advances for area-law entangled ground states~\cite{RevModPhys.93.045003}, the simulation of non-equilibrium states is generically still limited to short time due to the fast entanglement growth~\cite{paeckel2019time}, although in certain cases non-equilibrium phenomena are accessible even at long times using additional approximation~\cite{prosen2009matrix,cui2015variational,rakovszky2022dissipation}.

Recent developments of programmable quantum computers and simulators allow for large-scale studies of quantum many-body systems~\cite{Bluvstein_2021,daley2022practical}.
The simulation of non-equilibrium dynamics is one of the tasks where quantum advantage is anticipated in the near term, as its complexity scales linearly in the system size and time~\cite{wiebe2011simulating,childs2019nearly} on quantum devices.
In case of a finite system, several algorithms for current Noisy Intermediate Scale Quantum (NISQ) devices were developed to simulate quantum dynamics on a finite system~\cite{li2017efficient,Lin_2021,Barison_2021,https://doi.org/10.48550/arxiv.2204.03454,PhysRevResearch.3.033083,https://doi.org/10.48550/arxiv.2205.11427}.

{
Simulation of dynamics for formally infinite translationally invariant systems facilitates understanding of physics in the thermodynamic limit. However, the scaling of complexity for quantum algorithms in this limit is subtler than in the finite system case ~\cite{wiebe2011simulating,childs2019nearly}. Recent works have shown that is possible to simulate infinite systems with a finite number of qubits~\cite{Barratt_2021,dborin2022simulating}. It is now crucial to address the scalability and stability of quantum algorithms working in the thermodynamic limit. In this work, we present a hybrid quantum-classical algorithm for time-evolving translation invariant systems in one dimension and demonstrate both the expressibility and scalability of our algorithm.

%
%
%
%
}

We consider layered uniform sequential circuits ({\ourAnsatz}) ansatz, which is a generalization to the single-layer USC ansatz introduced in Ref.\,\cite{Barratt_2021}.
The ansatz {\ourAnsatz} forms a subclass of dense USC ({\denseAnsatz}), which are equivalent to matrix-product states~\cite{schon2005sequential,schon2007sequential}.
Moreover, we propose a gradient-based algorithm for time evolving quantum states within the manifold spanned by {\ourAnsatz}.
This includes a routine for computing the transfer matrix and environments of the uniform states that does not
require tomography or post-selection that would lead to an exponential scaling.

To benchmark the proposed algorithm, we simulate it on a classical computer. We show that the number of variational parameters required to accurately time-evolve a quantum state for the time $t$ with the {\ourAnsatz} ansatz, scales only polynomially in $t$.
Lastly, having obtained the time-evolved {\ourAnsatz} state representation on a classical computer, we compute physical observables on a cloud-based quantum processing unit (QPU) and demonstrate agreement with quasi-exact results obtained with { infinite time-evolution block-decimation (iTEBD) algorithm} at a large bond dimension~\cite{PhysRevLett.98.070201}.

This article is organized as follows. In Section~\ref{sec:method}, we introduce the layered uniform sequential circuit ansatz and the gradient-based variational time-evolution algorithm.
In Section~\ref{sec:results} we present the simulation results and analyze the effect of the layered decomposition on the accuracy of the time-evolved quantum state representation and fixed points of the transfer matrix. In addition, we show physical observables obtained from  the classically optimized circuit, measured on real quantum hardware.
In Section~\ref{sec:discussion} we discuss the obtained data and outline the prospects for future work.

\section{Methodology}
\label{sec:method}

In this section, we introduce {and provide motivation for} the {\ourAnsatz} ansatz, which is a subclass of {\denseAnsatz} where the dense unitary is replaced by the layered decomposition.
We present the necessary entities, e.\,g., transfer matrices and the environments, to measure the physical observables with {\ourAnsatz}.
We then turn to the algorithm for time-evolving the {\ourAnsatz} and the routines to perform the variational time evolution.

\subsection{The Layered Uniform Sequential Circuit Ansatz}

{
The main motivation for our ansatz stems from the great success of classical simulation for quantum systems with tensor network methods~\cite{cirac2021matrix}, especially with matrix-product states (MPS) applied to one-dimensional systems~\cite{verstraete2023density}.
The classical simulation methods are so efficient that it is argued that there might be no exponential quantum advantage with quantum algorithms for ground state problems in quantum chemistry~\cite{lee2023evaluating}.
In contrast, the fast growth of entanglement in quantum non-equilibrium dynamics makes classical tensor network methods inefficient due to the exponentially growing tensor size with respect to the evolution time.
However, for finite systems it has been shown that these tensors have a simple structure that can be efficiently represented as quantum-circuit ans\"atze~\cite{Lin_2021,Haghshenas_2022}.
In this work, we propose the {\ourAnsatz} for translationally invariant infinite systems.
}

\begin{figure}[t!]
    \centering
    \includegraphics[width=\columnwidth]{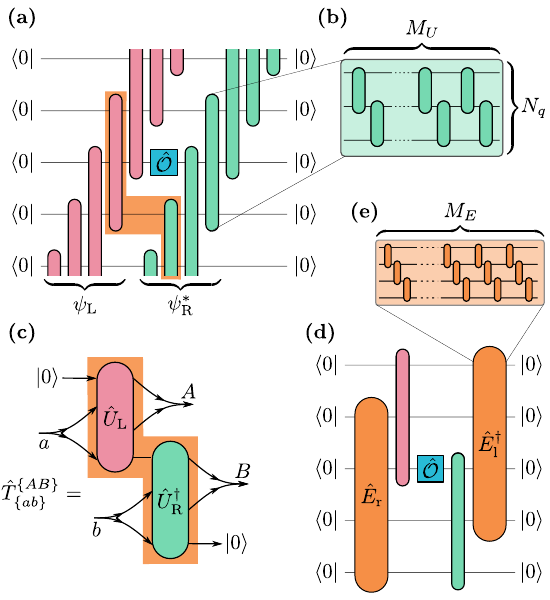}
    \caption{{\bf (a)}~Circuit representing $\langle \psi_{\text{R}}(\boldsymbol \theta) | \hat{\mathcal{O}} | \psi_{\text{L}}(\boldsymbol \theta') \rangle$ with $|\psi_{\text{L}}(\boldsymbol \theta')\rangle$ and $|\psi_{\text{R}}(\boldsymbol \theta) \rangle$ being the same state in left and right representations.
    The shaded region singles out the repeated circuit element, i.e., the transfer matrix.
    {\bf (b)}~Decomposition of a state-unitary into $M_U$ layers of sequential 2-qubit gates.
    {\bf (c)}~Transfer matrix $\hat{T}^{\{AB\}}_{\{ab\}}(\boldsymbol \theta, \boldsymbol \theta')$ between the left and right representations $|\psi_{\text{L}}(\boldsymbol \theta')\rangle$ and $|\psi_{\text{R}}(\boldsymbol \theta) \rangle$.
    {\bf (d)}~Circuit representation of $\langle l, 0 | \hat{U}_{\text{R}}^\dagger \hat{\mathcal{O}} \hat{U}_{\text{L}} | 0, r \rangle$ on finite number of qubits
    {\bf (e)}~Decomposition of an environment-unitary into $M_E$ layers of sequential 2-qubit gates.} 
    \label{fig:ansatz}
\end{figure}

{
An MPS can be equivalently represented as a sequential quantum circuit~\cite{schon2005sequential,schon2007sequential}, which we call a {\denseAnsatz}, shown in Fig.\,\ref{fig:ansatz}\,(a) for an infinite chain.
The {\denseAnsatz} define the wave functions}
\begin{equation}
    \label{eq:dense_ansatz}
    |\psi_{\text{R}}\rangle = \prod\limits_{i = -\infty}^{+\infty} \hat{U}^i_{\mbox{\footnotesize R}}(\boldsymbol \theta) |0 \rangle, \qquad
    |\psi_{\text{L}}\rangle = \prod\limits_{i = +\infty}^{-\infty} \hat{U}^i_{\mbox{\footnotesize L}}(\boldsymbol \theta') |0 \rangle ,
\end{equation}
where $\hat{U}^i_{\text{R}}(\boldsymbol \theta)$ and $\hat{U}^i_{\mbox{\footnotesize L}}(\boldsymbol \theta')$ are $i$--independent unitaries acting on $N_q$ consecutive qubits $i,\,i + 1,\,\ldots,\,i + N_q - 1$.
The `R' index denotes the {\it right representation} and similarly the {\it left representation} `L' is defined by a different unitary $\hat{U}_{\text{L}}(\boldsymbol \theta')$ acting in the opposite order.
We show in Appendix~\ref{appendix:allexact} that the {\denseAnsatz} ansatz in left and right representations over $N_q$ qubits are MPS in left and right isometric forms with the bond dimension $\chi = 2^{N_q - 1}$ respectively.

The {\ourAnsatz} ansatz is defined as a specific form of Eq.\,\eqref{eq:dense_ansatz}, where each unitary $\hat{U}_{\text{R}/\text{L}}$ is parameterized by a sequential circuit of $M_U$ layers, as shown in Fig.\,\ref{fig:ansatz}\,(b).
Each layer consists of a consecutive application of 2-qubit gates between neighboring qubits in the direction shown in Fig.~\ref{fig:ansatz}\,(a-b).
While for any {\denseAnsatz} state in the right representation, there exists an exact {\denseAnsatz} of the same size in the left representation, this does not always hold for {\ourAnsatz} with the same $N_q$ and $M_U$ unless the state is inversion symmetric.
We note that {\ourAnsatz} ansatz belongs to the broad class of quantum circuit tensor network ans\"atze~\cite{Haghshenas_2022}~\footnote{Following the naming scheme in~\cite{Haghshenas_2022}, our ansatz is the uniform qMPS-L. The direction of the application of layered $2$--site gates is, however, the opposite.}, where the dense unitaries in the isometric tensor networks are replaced by various kinds of local circuits, e.\,g., brick-wall circuits or sequential circuits.
{
The {\denseAnsatz} and {\ourAnsatz} wave functions are both universal:
if one allows arbitrary $N_q, M_U$, all translationally-invariant quantum many-body states can be approximated to arbitrary accuracy in either {\denseAnsatz} or {\ourAnsatz} form.
The required $N_q, M_U$ indicates the complexity of the quantum many-body state.
As an example, the Greenberger–Horne–Zeilinger (GHZ) state~\cite{greenberger1989going} can be represented exactly with $N_q=2, M_U=1$.
Rigorous studies of the scaling properties of the quantum circuit ansatz could give us more insight into the properties of quantum states, for example, the recent work for ground states~\cite{jobst2022finite}.
In this work, we focus on studying the expressivity of the ansatz applied to time-evolution with both a free-fermion and a non-integrable Hamiltonian.
}

The {\ourAnsatz} with local circuits acting on $N_q$ qubits defines a subclass of states within the manifold of {\denseAnsatz}, or equivalently uniform MPS of bond dimension $\chi=2^{N_q - 1}$.
As a generic 2-qubit gate, up to a global phase, requires $15$ parameters~\cite{https://doi.org/10.48550/arxiv.1905.13311}, the {\ourAnsatz} ansatz is parametrized by at most $15 (N_q - 1) M_U$ optimization parameters~\footnote{We note that the number of parameters can be reduced considering the redundancy of the consecutive single-qubit gates.}, as compared to $ 2^{2 N_q + 1}$ parameters necessary for the dense parametrization in the {\denseAnsatz} ansatz.
In the previous works, it has been shown that similar ans\"atze on a finite system are polynomially more efficient in representing ground states~\cite{Haghshenas_2022} and exponentially more efficient in representing time-evolved states~\cite{Lin_2021}.
{ Previous works studying the dynamics of infinite systems have been focused on the specific case $N_q=2, M_U=1$.
The question remains on whether there is also an exponential advantage in the thermodynamics limit in expressing quantum states produced under non-equilibrium dynamics.
}
Later in the Section~\ref{sec:results}, we will demonstrate that {\ourAnsatz} forms a physically relevant subset of the {\denseAnsatz} states with the corresponding bond dimension, and allows for efficient time-evolution of quantum states. We begin here with the description of the tools to acquire physical observable from the {\ourAnsatz} state representation.

\paragraph{Transfer matrix.}
Computation of physical observables and other operations of an infinite system can be performed on a finite number of qubits using the transfer matrix and its dominant eigenvectors, known as {\it environments} in the context of tensor networks.
Utilizing the left and right representation of {\ourAnsatz}, we always consider the (mixed) transfer matrix defined between the states in left representation, { $|\psi_{\text{L}}\rangle$}, and in right representation, { $|\psi_{\text{R}}\rangle$},  as shown as the shaded area in Fig.\,\ref{fig:ansatz}\,(a).
In Fig.\,\ref{fig:ansatz}\,(c), we explicitly write down the transfer matrix $\hat{T}^{\{AB\}}_{\{ab\}} (\boldsymbol \theta, \boldsymbol \theta')$, with $\{AB\}$ forming a united out-index and $\{ab\}$ forming a united in-index.
The arrow directions indicate the flow of time of the quantum circuit execution.

With this construction, the transfer matrix is a linear operator $T:V^{ab}\rightarrow V^{AB}$ mapping a pure state in Hilbert space $V^{ab}$ to a pure state in Hilbert space $V^{AB}$.
The linear map is realized by a combination of unitary operators  with the post-selection on one qubit, as shown in Fig\,\ref{fig:ansatz}\,(c). 
The transfer matrix is therefore generally non-Hermitian and non-unitary.
In Appendix\,\ref{appendix:post_selection_prob}, we show that the post-selection probability is close to unity for cases considered in this work.
This formalism comes from the construction of the transfer matrix using simultaneously left and right representations. This is different from Ref.\,\cite{Barratt_2021,dborin2022simulating}, where the transfer matrix is defined with the inner product of states in the same representation, and the transfer matrix is a quantum channel mapping between density matrices.

The left and right environments $|l \rangle$ and $|r \rangle$ are the dominant eigenvectors of the transfer matrix $\hat{T}$ satisfying the fixed point equations $\hat{T} |r\rangle = \lambda |r\rangle$, $\hat{T}^{\dagger} |l\rangle = \lambda^* |l\rangle$, where $\lambda$ is the eigenvalue of $\hat{T}$ with the maximum absolute magnitude.
The absolute value of the eigenvalue $\lvert \lambda \rvert \leqslant 1$ defines the overlap density between the two states, and $|\lambda| = 1$ if and only if the states are identical. In such case, the left and right environments are identical up to complex conjugation, as we prove in Appendix\,\ref{appendix:identical_fixed_points}.

From the construction of the transfer matrix, these environments are of dimension $2^{2 N_q - 2}$.
To translate the environments into variational quantum circuits, we introduce two $2^{2 N_q - 2} \times 2^{2 N_q - 2}$ parametrized environment unitaries $\hat{E}_{\text{r}}$ and $\hat{E}_{\text{l}}$, such that $|r\rangle = \hat{E}_{\text{r}}(\boldsymbol{\varphi}_{\text{r}}) |0\rangle$ and $|l\rangle = \hat{E}_{\text{l}}(\boldsymbol{\varphi}_{\text{l}}) |0\rangle$, as shown in Fig.\,\ref{fig:ansatz}\,(d).
Ultimately, we also consider the decomposition of environment unitaries in the form of the sequential circuits decomposition with $M_E$ layers, as shown in Fig.\,\ref{fig:ansatz}\,(e).
We discuss the method of obtaining the environments in the next section.

\paragraph{Evaluating local observables.}
We evaluate the expectation value of an local observables utilizing the mixed representation,
\begin{equation}
  \langle \hat{\mathcal{O}}\rangle = \frac{\langle \psi | \hat{\mathcal{O}} | \psi \rangle}{\langle \psi |  \psi \rangle}  
  = \frac{\langle \psi_{\text{R}}(\boldsymbol \theta) | \hat{\mathcal{O}} | \psi_{\text{L}}(\boldsymbol \theta') \rangle}{\langle \psi_{\text{R}}(\boldsymbol \theta) | \psi_{\text{L}}(\boldsymbol \theta') \rangle} .
\end{equation}
In Fig.\,\ref{fig:ansatz}\,(a), we show the circuit representation of  the numerator $ \langle \psi_{\text{R}}(\boldsymbol \theta) | \hat{\mathcal{O}} | \psi_{\text{L}}(\boldsymbol \theta') \rangle$, where $\hat{\mathcal{O}}$ is a local observable that is Hermitian and unitary.
Using the definition of the environments, the expectation reduces to
\begin{equation}
  \langle \hat{\mathcal{O}}\rangle = \frac{\langle l, 0 | \hat{U}_{\text{R}}^\dagger \hat{\mathcal{O}} \hat{U}_{\text{L}} | 0, r \rangle}{ \langle l, 0 | \hat{U}_{\text{R}}^\dagger \hat{U}_{\text{L}} | 0, r \rangle} = \frac{\langle l, 0 | \hat{U}_{\text{R}}^\dagger \hat{\mathcal{O}} \hat{U}_{\text{L}} | 0, r \rangle}{\lambda \langle l | r \rangle}.
\end{equation}
Therefore, the expectation value of local observables can be evaluated by measuring finite circuits, which can be implemented on a quantum computer. 
The projective measurement on $|00\ldots0\rangle$ at the end of the circuit in Fig.\,\ref{fig:ansatz}\,(d) has the probability equal to the squared magnitude of the expectation value $\lvert \langle l, 0 | \hat{U}_{\text{R}}^\dagger \hat{\mathcal{O}} \hat{U}_{\text{L}} | 0, r \rangle \rvert^2$.
The same applies for the denominator.
Combining this together, one can measure the squared magnitude of the expectation value $|\langle \hat{\mathcal{O}}\rangle|^2$.
In Appendix~\ref{appendix:TM}, we provide the derivation of the above equations.
In the next section, we will describe the procedure to measure the expectation $\langle \hat{\mathcal{O}}\rangle$, including both real and imaginary parts.

We note that the outlined procedure can be generalized to evaluating correlation functions of the form $\langle \psi| \hat{A}_i \hat{B}_{i + \delta} | \psi \rangle$, where the operators $\hat{A}_i$, $\hat{B}_{i + \delta}$ act on single qubits and are separated by $\delta$ sites.

\subsection{Translationally-invariant Trotterization}

The time evolution of an initial wave function $|\psi_0 \rangle$ under the action of a Hamiltonian $\hat{H}$ is given by application of the evolution operator to the initial state $|\psi(t) \rangle = \hat{U}_t |\psi_0 \rangle = \exp (-i t \hat {H}) |\psi_0 \rangle$. 
Here, we consider a Hamiltonian acting on a one-dimensional infinite spin--$1/2$ chain. When $\hat{H}$ is local, i.\,e., can be written as $\hat{H} = \sum_i \hat{h}_i$ with all terms $\hat{h}_i$ having a finite support, we can approximate the evolution operator $\hat{U}_T$ using a sequential Trotter decomposition.
A first-order sequential Trotterization can be written as
\begin{gather}
\label{eq:TS_Trotter}
    \hat{U}_T = \left(\hat{u}(\delta t) \right)^k + \mathcal{O}(k \delta t^2),
\end{gather}
where $\hat{u}(\delta t) = \prod_j \hat{u}_j(\delta t)$, $\hat{u}_j(\delta t) = \exp ( i \delta t \hat{h}_j)$ and $\delta t = T / k$. A single sequential evolution operator $\hat{u}(\delta t)$ is shown in Fig.\,\ref{fig:Tevolution}\,(a). 
Due to the sequential decomposition, $\hat{u}(\delta t)$ and hence $\hat{U}_T$ are translationally invariant with a single site unit cell.
Starting with {a translationally} invariant state, we always need only a single unitary $\hat{U}(\boldsymbol \theta)$ parameterizing the state as in Eq.~\eqref{eq:dense_ansatz}~\footnote{This is in contrast to the classical iTEBD algorithm, where the evolution unitaries at even and odd chain sites are applied consequently, which results in two-site unit cell~\cite{PhysRevLett.98.070201,Or_s_2008}.}. All these considerations can be generalized to cases with a larger unit cell.

\subsection{The time evolution algorithm}

\begin{figure*}[t!]
    \centering
    \includegraphics[width=\textwidth]{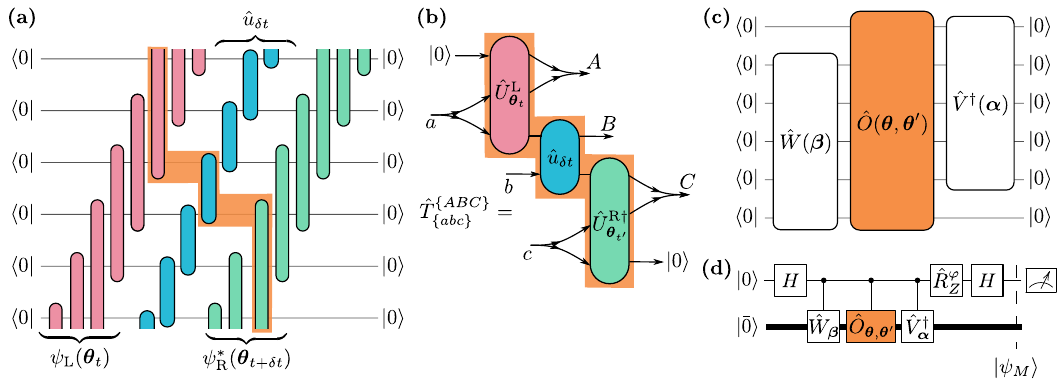}
    \caption{
    {\bf (a)}~Quantum circuit for the overlap between the time-evolved left-represented state $\hat{u}(\delta t) | \psi_{\text{L}}(\boldsymbol \theta_{t}) \rangle$ at time $t$ and right-represented state $|\psi_{\text{R}}(\boldsymbol \theta_{t + \delta t}) \rangle$ at time $t + \delta t$.
    For illustration, here $N_q = 3$ is chosen. The orange shaded area singles out the transfer matrix.
    {\bf (b)}~The explicit form of the transfer matrix $\hat{T}^{\{ABC\}}_{\{abc\}}(\boldsymbol \theta_t, \boldsymbol \theta_{t + \delta t})$. The capital indices $\{ABC\}$ form a composite out-index, similarly $\{abc\}$ form a composite in-index.
    The arrow direction indicates the operation order (time flow).
    {\bf (c)}~The circuit of the generalized functional $\mathcal{L}$.
    The unitary $\hat{W} (\boldsymbol \beta)$ acts on the first $2 N_q - 1$ qubits of the $|00\ldots0\rangle$ state prepared on $2 N_q$ qubits, then unitary $\hat{O}(\boldsymbol \theta, \boldsymbol \theta')$ acts all qubits and $\hat{V}^{\dag} (\boldsymbol \alpha)$ acts on the last $2 N_q - 1$ qubits.
    {\bf (d)} General Hadamard test scheme for measurement of the algebraic value of $\mathcal{L}$.
    }
    \label{fig:Tevolution}
\end{figure*}

We now introduce a hybrid quantum-classical algorithm to the perform time evolution of the {\ourAnsatz} representation.
At the time $t$, we parametrize the state-unitary $\hat{U}_{\text{R}/\text{L}}(t)$ by a set of variational parameters $\boldsymbol \theta_t$. The gradients of the parameters are measured on a quantum computer and the update is performed on a classical computer.
Here, for the sake of concrete notation, we present the {\it even} time steps of the algorithm. In these steps, representation of the wave function flips from left to right. The {\it odd} steps are done similarly, but with flipping from right representation to left.

To perform the time evolution at an even step, one is required to find the closest state $|\psi_{\text{R}}(\boldsymbol \theta_{t + \delta t}) \rangle$ in right representation approximating the time-evolved state $\hat{u}(\delta t) |\psi_{\text{L}}(\boldsymbol \theta_t) \rangle$.
The direct measure of the closeness is the fidelity, i.\,e., squared overlap, between the two states,   
\begin{equation*}
    \lvert \xi(t, \delta t) \rvert^2 = \lvert \langle \psi_{\text{R}}(\boldsymbol \theta_{t + \delta t}) | \hat{u}(\delta t) | \psi_{\text{L}}(\boldsymbol \theta_{t}) \rangle \rvert^2 .
\end{equation*}
It is the probability of measuring the state $|\ldots000\ldots\rangle$ at the end of the circuit shown in Fig.~\,\ref{fig:Tevolution}\,(a).
This quantity is either $1$ or $0$ in the thermodynamic limit and cannot be used for posing the optimization problem.
Instead, we define the mixed transfer matrix between the two states over indices $\{abc\},\,\{ABC\}$ as shown in Fig.\,\ref{fig:Tevolution}\,(a-b) with one additional index coming from the trotterized unitary.
To find the closest state, we maximize the absolute value of the overlap density $\lvert \lambda \rvert$ with respect to the parameters at the next time step $\boldsymbol \theta_{t + \delta t}$.
The squared magnitude of the overlap density $|\lambda|^2$ is the probability of measuring the state $|00\ldots0\rangle$ at the end of the circuit shown in Fig.\,\ref{fig:Tevolution}\,(c).

We solve the maximization problem with a gradient ascent algorithm which requires the knowledge of environments $|l\rangle$ and $|r\rangle$ and the leading eigenvalue from the mixed transfer matrix.
To obtain the environments $|l\rangle$ and $|r\rangle$, we employ the {\it modified power method}.
We describe the procedure of obtaining the right environment $|r\rangle$, while the procedure for the left environment is similar, apart from the replacement $T \to T^{\dagger}$.
The idea of the power method is to take an initial state $|\psi_0\rangle$ and project it onto the leading eigenvector of $\hat{T}$ by repeated application of $\hat{T}$, because $\lim_{p\rightarrow \infty} (T/\lambda)^p \rightarrow |r\rangle \langle l|$.
Here, we consider an iterative algorithm, which is a slight modification of the power method: at each step, we find the new vector $|r(\boldsymbol \varphi_{\text{r}}')\rangle$ by performing only a single gradient descent step maximizing the overlap magnitude $|\lambda|^2 = \lvert \langle r(\boldsymbol \varphi_{\text{r}}') | \hat{T} |r(\boldsymbol \varphi_{\text{r}})\rangle \rvert^2$ with respect to $\boldsymbol \varphi_{\text{r}}'$.
Namely, $\boldsymbol \varphi_{\text{r}}' \leftarrow \boldsymbol \varphi_{\text{r}}' + \eta \boldsymbol{\nabla}_{\boldsymbol \varphi_{\text{r}}'} |\lambda|^2$, where $\eta$ is the learning rate.
Alternatively, a gradient-free method, such as Rotosolve~\cite{vidal2018calculus,nakanishi2020sequential,parrish2019jacobi,ostaszewski2021structure}, could be used.
The environment  vector is then updated $|r(\boldsymbol \varphi_{\text{r}})\rangle \leftarrow |r(\boldsymbol \varphi_{\text{r}}')\rangle$ and is used in the next iteration.
At each step, $|r(\boldsymbol \varphi_{\text{r}})\rangle$ has a strictly increasing overlap with the leading eigenvector of $\hat{T}$ provided a small enough step size $\eta$. The method is presented in Algorithm~\ref{alg:power_method}.

\begin{algorithm}[H]
\caption{The power method for an environment}\label{alg:power_method}
\begin{algorithmic}[1]
\Procedure{Environment}{$\hat{T}, \boldsymbol \varphi_{\text{r}}^p$}
    \State $\boldsymbol \varphi_{\text{r}}, \boldsymbol \varphi_{\text{r}}' \leftarrow \boldsymbol \varphi_{\text{r}}^p$  \Comment{from the previous step}
    \While{not converged}
        \State \text{Measure} $\lambda = \langle 0 | \hat{E}^{\dagger}_{\text{r}}(\boldsymbol \varphi_{\text{r}}') \hat{T} \hat{E}_{\text{r}}(\boldsymbol \varphi_{\text{r}})| 0 \rangle$
        \State \text{Measure} $\boldsymbol{\nabla}_{\boldsymbol \varphi_{\text{r}}'} \lambda$ \Comment{See Eq.\,\eqref{eq:der_T}}
        \State $\boldsymbol \nabla_{\boldsymbol \varphi_{\text{r}}'} |\lambda|^2 = 2 \mbox{Re}\, \left[ \lambda^* \boldsymbol \nabla_{\boldsymbol \varphi_{\text{r}}'} \lambda \right]$
        \State $\boldsymbol \varphi_{\text{r}}' \leftarrow \boldsymbol \varphi_{\text{r}}' + \eta \boldsymbol \nabla_{\boldsymbol \varphi_{\text{r}}'} |\lambda|^2$
        \State $\boldsymbol \varphi_{\text{r}} \leftarrow \boldsymbol \varphi_{\text{r}}'$
    \EndWhile
    \State \textbf{return} $|r(\boldsymbol \varphi_{\text{r}})\rangle = \hat{E}_{\text{r}}(\boldsymbol \varphi_{\text{r}}) |\vec{0}\rangle$
\EndProcedure
\end{algorithmic}
\end{algorithm}

Next, we show in Algorithm\,\ref{alg:te_algorithm} how to perform a time evolution step using gradient ascent  method with the environments we obtained.
The algorithm uses a nested variational approach, in which left- and right-environments are variationally optimized between the consecutive gradient descent steps.
Both algorithms are run until the change of $|\lambda|^2$ between two consecutive iterations becomes smaller than $10^{-12}.$

\begin{algorithm}[H]
\caption{Time evolution algorithm for {\ourAnsatz}}
\label{alg:te_algorithm}
\begin{algorithmic}[1]

\Procedure{Evolution step}{$\boldsymbol \theta_t, \boldsymbol \varphi_{\text{r}}^t, \boldsymbol \varphi_{\text{l}}^t$}
    \State $\boldsymbol \theta_{t + \delta t}
    \leftarrow \boldsymbol \theta_t$  
    \State $\boldsymbol \varphi_{\text{r}}, \boldsymbol \varphi_{\text{l}} \leftarrow
    \boldsymbol \varphi_{\text{r}}^t, \boldsymbol \varphi_{\text{l}}^t$

    \While{not converged}
        \State $\boldsymbol \varphi_{\text{r}} \leftarrow \text{Environment}(\hat{T}(\boldsymbol \theta_t, \boldsymbol \theta_{t + \delta t}), \boldsymbol \varphi_{\text{r}})$;
        \State $\boldsymbol \varphi_{\text{l}} \leftarrow \text{Environment}(\hat{T}^{\dagger}(\boldsymbol \theta_t, \boldsymbol \theta_{t + \delta t}), \boldsymbol \varphi_{\text{l}})$;
        \State $|r\rangle \leftarrow \hat{E}_{\text{r}}(\boldsymbol \varphi_{\text{r}})$, $|l\rangle \leftarrow \hat{E}_{\text{l}}(\boldsymbol \varphi_{\text{l}})$;
        \State $\lambda = \langle l | \hat{T}(\boldsymbol \theta_t, \boldsymbol \theta_{t + \delta t}) | r \rangle / \langle l | r \rangle$;
        \State $\displaystyle \boldsymbol \nabla_{\boldsymbol \theta_{t + \delta t}} |\lambda|^2 = 2 \,\mbox{Re} \left[ \lambda^* \boldsymbol \nabla_{\boldsymbol \theta_{t + \delta t}} \lambda \right]$; \Comment{See Eq.\,\eqref{eq:der_T}}
        \State $\boldsymbol \theta_{t + \delta t} \leftarrow \boldsymbol \theta_{t + \delta t} + \eta \boldsymbol \nabla_{\boldsymbol \theta_{t + \delta t}} |\lambda|^2$;
    \EndWhile
    \State {\bf return} $\boldsymbol \theta_{t + \delta t}$
\EndProcedure

\end{algorithmic}
\end{algorithm}

Note that we can use the algorithm to find the opposite representation of the same wave function if the time evolution operator is taken to be the  identity.
The algorithm proposed here resembles the time evolution algorithm for a finite size system~\cite{otten2019noise,Lin_2021} and for an infinite system~\cite{Barratt_2021}. As the main difference, in this work the (mixed) transfer matrix is constructed as the mixed representation $\langle \psi_{\text{R}}(\boldsymbol{\theta}) |  \psi_{\text{L}}(\boldsymbol{\theta}') \rangle$. The environments of the transfer matrix are represented as quantum states parametrized with layered sequential circuits. We study the effect of such approximation in Sec.\,\ref{sec:results}\,(B).

\subsection{Required measurements}
To implement the time-evolution Algorithm\,\ref{alg:te_algorithm} in practice, one has to measure the algebraic value of the overlap $\lambda = \langle 0 | \hat{E}_{\text{r}}^{\dagger}(\boldsymbol \varphi_{\text{r}}') \hat{T} \hat{E}_{\text{r}}(\boldsymbol \varphi_{\text{r}})| 0 \rangle$, of its derivative with respect to the parameters of the environment $\boldsymbol \nabla_{\boldsymbol\varphi_{\text{r}}'} \lambda = \langle 0 | \boldsymbol \nabla_{\boldsymbol\varphi_{\text{r}}'} \hat{E}_{\text{r}}^{\dagger}(\boldsymbol \varphi_{\text{r}}') \hat{T} \boldsymbol \hat{E}_{\text{r}}(\boldsymbol \varphi_{\text{r}})| 0 \rangle$, and of its derivative with respect to the state unitary 
\begin{gather}
    \label{eq:der_T}
    \boldsymbol \nabla_{\boldsymbol \theta_{t + \delta t}} \lambda = \frac{\langle l | \boldsymbol \nabla_{\boldsymbol \theta_{t + \delta t}} \hat{T}(\boldsymbol \theta_t, \boldsymbol \theta_{t + \delta t})|r \rangle}{\langle l | r \rangle}.
\end{gather}
We prove the latter formula in Appendix\,\ref{appendix:TM_derivative}. Notably, the implicit dependency of the right- and left-environments on $\boldsymbol \theta_{t + \delta t}$ gives no contribution to the gradient.
These expectation values can all be expressed in terms of a general overlap functional $\mathcal{L}[\hat{V} (\boldsymbol \alpha), \hat{O}(\boldsymbol \theta, \boldsymbol \theta'), \hat{W} (\boldsymbol \beta)] = \langle 0| \hat{V}^{\dag} (\boldsymbol \alpha) \hat{O}(\boldsymbol \theta, \boldsymbol \theta') \hat{W} (\boldsymbol \beta) |0\rangle$. In our case, $\hat{V} (\boldsymbol \alpha)$ and $\hat{W} (\boldsymbol \beta)$ are the environment unitaries or their derivatives, while $\hat{O}(\boldsymbol \theta, \boldsymbol \theta')$ is the transfer matrix or its derivatives. We note that all mentioned unitaries' derivatives are also unitary due to the specific parametrization of the two-qubit gates (for details, see Appendix\,\ref{appendix:classical_optimization}).

Absolute and algebraic values of this functional can be measured on a quantum computer. First, the square of the magnitude $|\mathcal{L}|^2$ is given by the probability of projection onto the $|00\ldots0\rangle$ state in the circuit in Fig.\,\ref{fig:Tevolution}\,(c)~\cite{Lim_2006}.
The algebraic value of the expectation $\mathcal{L}$ and its derivative  $\boldsymbol \nabla_{\boldsymbol\varphi_{\text{r}}'} \mathcal{L}$ can be measured within the Hadamard test procedure~\cite{Mitarai_2019} shown in Fig.\,\ref{fig:Tevolution}\,(d).
We denote $|\bar{0}\rangle = |00\ldots0\rangle$ and the quantum state before the ancilla qubit measurement reads
\begin{align}
    |\psi_M\rangle = &\frac{1}{2} |0 \rangle \otimes \left(|\bar{0}\rangle + e^{i \varphi} \hat{V}^{\dag} (\boldsymbol \alpha) \hat{O}(\boldsymbol \theta, \boldsymbol \theta') \hat{W} (\boldsymbol \beta) |\bar{0}\rangle\right) \nonumber\\
    &\ \ +\frac{1}{2} |1 \rangle \otimes \left(|\bar{0}\rangle - e^{i \varphi} \hat{V}^{\dag} (\boldsymbol \alpha) \hat{O}(\boldsymbol \theta, \boldsymbol \theta') \hat{W} (\boldsymbol \beta) |\bar{0}\rangle \right) .
\end{align}
The probability difference in measurement over the ancilla qubit yields 
\begin{equation}
p(|0\rangle) - p(|1\rangle) = \mbox{Re}\left[ e^{i \varphi} \langle \bar{0}|\hat{V}^{\dag} (\boldsymbol \alpha) \hat{O}(\boldsymbol \theta, \boldsymbol \theta') \hat{W} (\boldsymbol \beta) |\bar{0}\rangle \right].
\end{equation}
This scheme can also be used for obtaining algebraic values of the observable expectation $\langle \mathcal{O} \rangle$ introduced in Section II.\,A.
%

\begin{figure*}[t!]
    \centering
    \includegraphics[width=\textwidth]{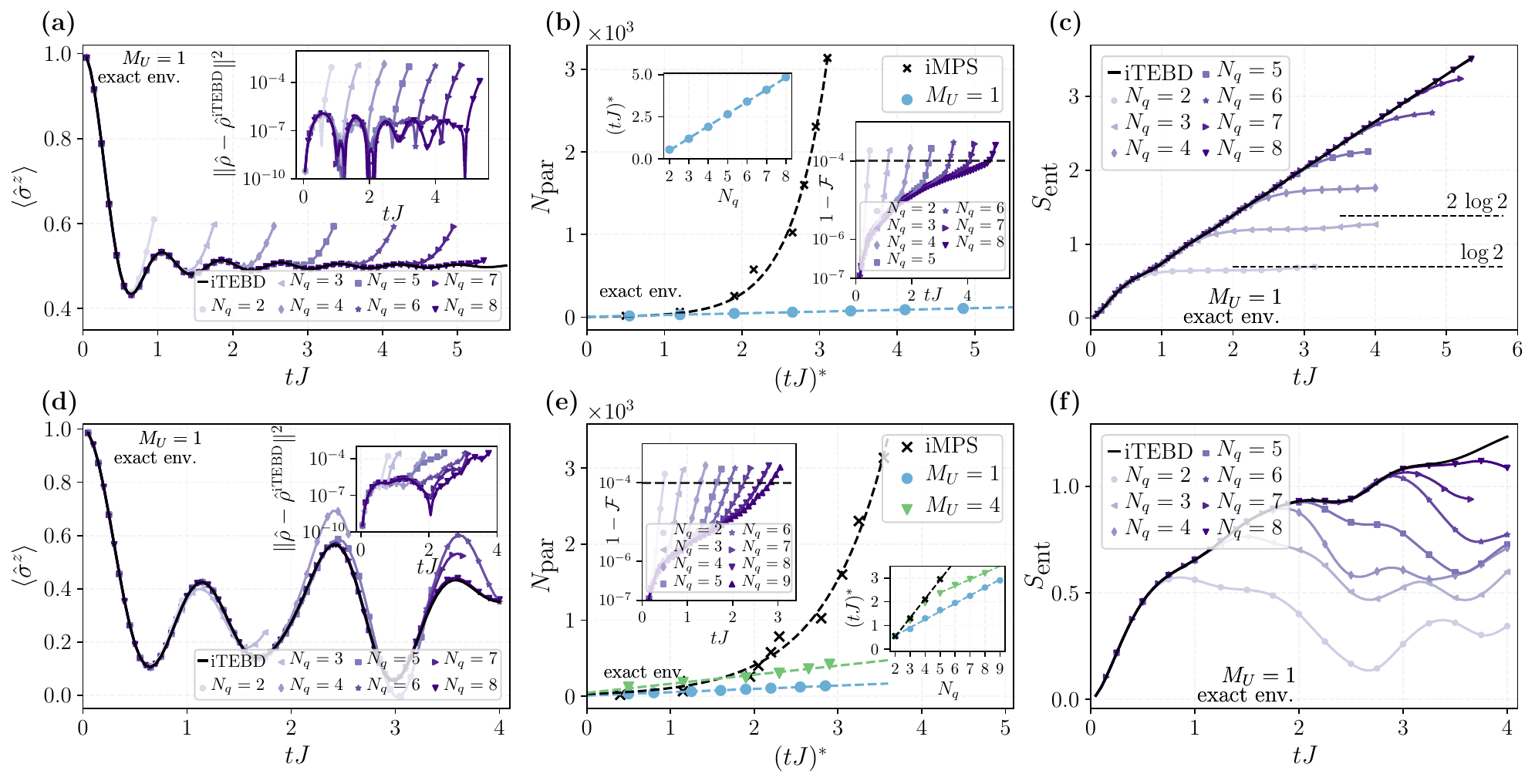}
    \caption{
     Simulations of time evolution using {\ourAnsatz} with $M_U = 1$ and exact environment for the Hamiltonian in Eq.\,\eqref{eq:hamiltonian} with $g / J = 1.0$, $h / J = 0$ { (upper row) and $h / J = 1$ with $M_U = 1$ and $4$ (lower row)}.
    {\bf (a)} Expectation value of $\langle \hat{\sigma}^z(t)\rangle$ using {\ourAnsatz} with different values of $N_q$.
    Inset: the difference in the single-site density matrices between the quasi-exact iTEBD simulation and optimization of {\ourAnsatz}.
    {\bf (b)} The number of parameters required to reach time $t^*$ with the fidelity density at least $\mathcal{F} = 1 - 10^{-4}$.
    The black crosses represent the standard iTEBD approach with the black dashed line showing an exponential fit.
    The blue markers show the results for {\ourAnsatz}, while the line shows the linear fit.
    Left inset: the reachable time $t^*$ under the condition $\mathcal{F} \geqslant 1 - 10^{-4}$ as a function of $N_q$. The dashed line shows a linear fit.
    Right inset: error in fidelity density $1-\mathcal{F}$ for different $N_q$.
    {\bf (c)} Entanglement entropy as a function of evolution time $t J$ compared with the quasi-exact result.
    The horizontal dashed lines mark the theoretical maximum entanglement entropy levels $(N_q - 1) \log 2$.
    {{\bf (d-f)} Same results for the $g / J = h / J = 1$ case. In the right inset of {\bf(e)}, linear fits start at $N_q = 2$ and $5$ for $M_U = 1$ and $4$, respectively.}
    \label{fig:panel_results_1}
    }
\end{figure*}

\section{Results}
\label{sec:results}

To benchmark the proposed algorithm, we simulate the quenched dynamics of the transverse-field Ising model with the longitudinal field
\begin{gather}
    \label{eq:hamiltonian}
    \hat{H} = J \sum\limits_{i} \sigma^x_i \sigma^x_{i + 1} + g \sum\limits_i \sigma^z_i + h \sum\limits_i \sigma^x_i
\end{gather}
over an infinite spin--$1/2$ chain~\footnote{{ Generalization to a Hamiltonian with longer-range interaction acting between qubits separated by $d$ would require a transfer matrix in Fig.\,\ref{fig:Tevolution} act on $2 N_q - 2 + d$ qubits.}}.
The initial wave function is taken as a fully-magnetized state $|\psi_0 \rangle = |\ldots 000\ldots\rangle$ in the $\sigma^z$ basis.
We use a fourth-order Trotterized iTEBD simulation with $\chi = 1024$ uniform MPS and $\delta t J = 10^{-2}$ as a quasi-exact reference labeled {\it iTEBD} in all figures.
In the following, we simulate the algorithm on a classical computer to study the properties of the {\ourAnsatz} ansatz.
The complexity of the {\ourAnsatz} ansatz is controlled by $M_U$ and $N_q$. The complexity of measuring local observables and running the time-evolution algorithm depends additionally on $M_E$, i.\,e., the accuracy of approximating the environments.
In Sec.\,\ref{sec:result_1}, we study the effect of varying $N_q$ and $M_U$ in simulating the time evolution with exact environment obtained by exact diagonalization of the transfer matrix.
In Sec.\,\ref{sec:result_2}, we study the accuracy of approximating an exact environment with a layered circuit with finite $M_E$, derive the relation between the required $M_E$ and $N_q$ and perform the full realistic simulation with both state and environment presented in the sequential form. Finally, in Sec.\,\ref{sec:result_3}, we demonstrate measuring the evolution of physical observables on a QPU with a classically optimized {\ourAnsatz} circuit.

\subsection{Layered state unitary, exact environment}
\label{sec:result_1}
As the first step, we study the performance of the {\ourAnsatz} ansatz using exact environments obtained through direct diagonalization of the transfer matrix, and perform the algorithm outlined in Algorithm~\ref{alg:te_algorithm}.
In Fig.\,\ref{fig:panel_results_1}, we show the simulation results obtained with $M_U = 1$ and various $N_q$.

In Fig.\,\ref{fig:panel_results_1}\,(a), we plot the evolution of the local magnetization $\langle \sigma^z(t) \rangle $ and observe that the time of deviation from the quasi-exact solution increases with $N_q$.
In the inset, we plot the Frobenius norm squared of the difference in the single-site density matrices between the quasi-exact state and the {\ourAnsatz}, i.\,e. $\lVert \rho - \rho^{\text{exact}} \rVert^2$.
The difference shows fluctuating behavior as a function of $t J$, but at some point shows rapid growth.
This fast growth coincides in evolution time $t$ with the noticeable discrepancy in $\langle \sigma^z(t) \rangle $.

To quantify the representation capacity of {\ourAnsatz}, we define the reachable time $t^*$ of the given ansatz as the time when the error in fidelity density with the quasi-exact (iTEBD) state crosses the threshold value $1-\mathcal{F} = 10^{-4}$. Here, $\mathcal{F}$ is the fidelity density, i.\,e., squared overlap per unit cell, between the {\ourAnsatz} state and the iTEBD wave function.
In Fig.\,\ref{fig:panel_results_1}\,(b), we plot the number of parameters in a given circuit against the reachable dimensionless time $t^*J$.
We see that within {\ourAnsatz}, the required number of parameters grows linearly with the reachable time.
Note that, in contrast, the number of parameters required to store a {\denseAnsatz} state grows exponentially in the reachable time $t^*J$.
Therefore, the {\ourAnsatz} ansatz defines a sub-manifold of uniform MPS that is relevant for representing states under time evolution. Namely, the {\ourAnsatz} is ``sparse'' as compared to {\denseAnsatz} (uniform MPS) and requires exponentially fewer parameters.

In practice, one does not have access to the exact state and, therefore, no access to the error in fidelity.
Instead, one can utilize the leading transfer matrix eigenvalues $\{\lambda_i\}$ obtained at all steps of the time evolution and define the accumulated error measure $\mathcal{M}(t) = 1 - \prod_{i < t} |\lambda_{i}|^2$ to monitor the error and understand whether the simulation result is reliable. 
In Appendix\,\ref{appendix:M_measure}, we demonstrate that this measure follows closely the true infidelity $1 - \mathcal{F}$ and thus can be used for assessment of the optimization quality. 

\begin{figure}[t!]
    \centering
    \includegraphics[width=1.0\columnwidth]{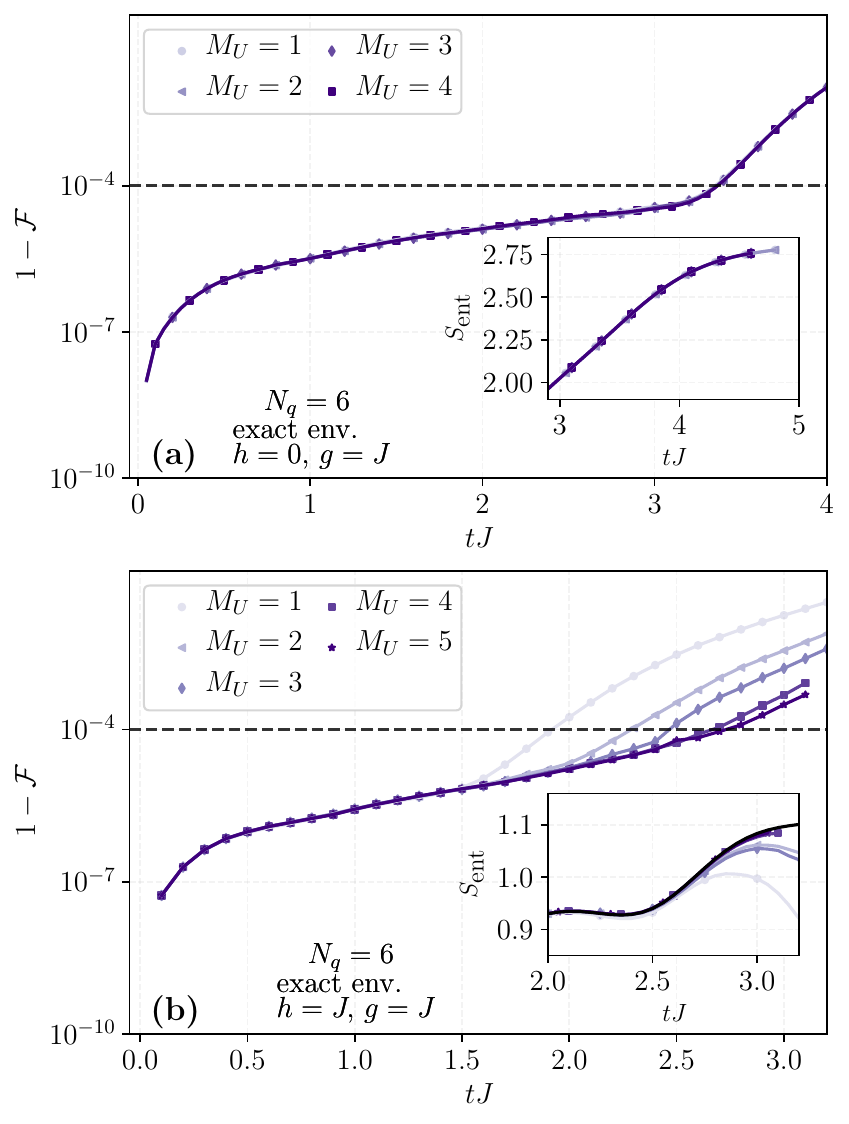}
    \caption{Fidelity density between the quasi-exact simulation and the time-evolved {\ourAnsatz} state with $N_q = 6$ and varied number of layers $M_U$.
    Inset: the entanglement entropies obtained within optimization of the {\ourAnsatz} circuit at $N_q = 6$. { {\bf (a)} The integrable case $h = 0$. {\bf (b)} Non-integrable case $h = J$.}}
    \label{fig:MU_variation}
\end{figure}

Lastly, in Fig.\,\ref{fig:panel_results_1}\,(c) we show entanglement entropy as a function of $t J$.
The deterioration of wave function quality, as shown in Fig.\,\ref{fig:panel_results_1}\,(a-b), is clearly connected to saturation of capability of a circuit with given $N_q$ to encode the linearly-growing entanglement entropy of the system.
An {\ourAnsatz} of given $N_q$ could encode at most $S_{\text{ent.}} \leqslant (N_q - 1) \log 2$ entanglement entropy.
We show in Fig.\,\ref{fig:panel_results_1}\,(c) that the entanglement grows linearly with time up to saturation.
We observe that for $N_q = 2$ and $3$ the saturated entanglement entropy is close to the theoretical bound and for $N_q \geqslant 4$ the entanglement entropy does not reach the theoretical bound.

To study the effect of increasing the $M_U$ in the layered sequential unitary decomposition.
Strikingly, in the integrable case $h = 0$ considered in Fig.\,\ref{fig:MU_variation}\,(a), we observe that increasing $M_U$ leads to negligible improvement in the reachable time $t^* J$, as compared with the effect of $N_q$. Similarly, increasing $M_U$ does not lead to significant change in the entanglement entropy of the time-evolved {\ourAnsatz} ansatz. 
In Appendix\,\ref{appendix:MU_effect}, we numerically demonstrate that in the integrable $h = 0$ case, time-evolution of the {\ourAnsatz} ansatz at $M_U = 1$ leads to the same wave function accuracy, as optimization of the full dense {\denseAnsatz} ansatz { of the same $N_q$.}
%
{ In Appendix\,\ref{appendix:sufficiency_proof}, we explain why for the case of the free-fermion model, $h = 0$, the $M_U = 1$ {\ourAnsatz} ansatz is sufficient and is equivalent to any higher $M_U$ {\ourAnsatz} ansatz. In summary, the Gaussian $M_U = 1$ {\ourAnsatz} ansatz is equivalent to Gaussian $N_q$ {\denseAnsatz} circuit, when the quantum Yang–Baxter equation is satisfied.}

\begin{figure}[t!]
    \centering
    \includegraphics[width=\columnwidth]{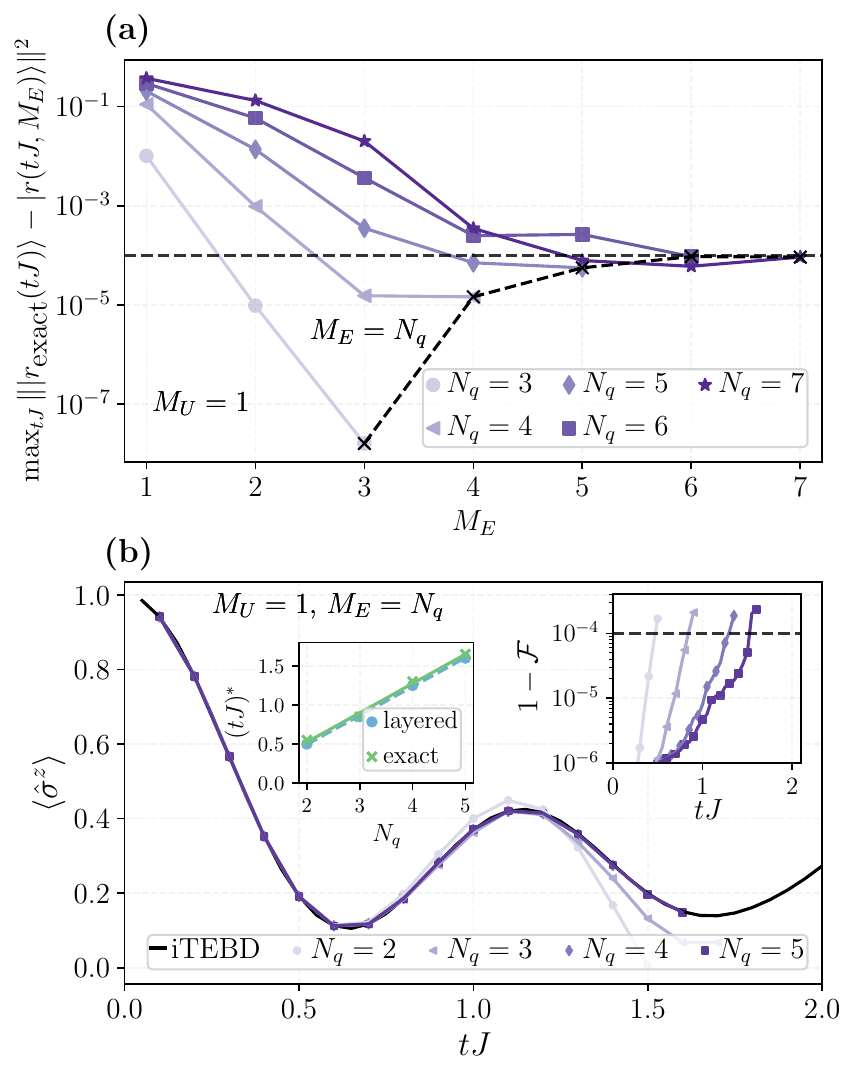}
    \caption{{\bf (a)} Maximum (over evolution time) error of approximating the exact environment by the $M_E$--layered environment throughout the full simulation shown in Fig.\,\ref{fig:panel_results_1}\,(b). { Here, we consider $h = g = J$.}
    The black dashed line shows the environment approximation error at $M_E = N_q$.
    {\bf (b)} Expectation value $\langle \hat{\sigma}^z(t)\rangle$ obtained within the full algorithm of time-evolution of {\ourAnsatz} with $M_U = 1$ and layered environment with $M_E = N_q$.
    Left inset: the reachable time $t^*$ of simulations with the exact environment and the layered environment with $M_E = N_q$. The lines show linear fits.
    Right inset: error in fidelity density $1-\mathcal{F}$ for different $N_q$.
    }
    \label{fig:max_E_err}
\end{figure}

{ The picture changes significantly in the non-integrable case, $h / J = 1$. In Fig.\,\ref{fig:panel_results_1}\,(d) we plot the evolution of the local magnetization, where we observe that the time of deviation from the quasi-exact solution increases with $N_q$, similarly to Fig.\,\ref{fig:panel_results_1}\,(a).
In Fig.\,\ref{fig:panel_results_1}\,(e), we show the linear growth of the number of parameters in a circuit against the reachable time $t^* J$, for both the $M_U = 1$ and $M_U = 4$ {\ourAnsatz} wave functions. In this non-integrable case, we observe that the slope depends on $M_U$. In the right inset, we plot reachable time as a function of $N_q$ for the {\denseAnsatz} ansatz (MPS) and the {\ourAnsatz} ansätze. As expected, at $N_q = 2$ all reachable times coincide. Then, at large $N_q \gtrsim 5$, the {\ourAnsatz} reachable times grow linearly with $N_q$. At small $N_q$, the $M_U = 4$ {\ourAnsatz} reachable times coincide with the ones of {\denseAnsatz}, since large $M_U$ allows us to approximate a generic dense $N_q$ unitary with large precision. 
The left inset shows the infidelity density as a function of $t J$ for the $M_U = 1$ {\ourAnsatz} ansatz.
Notably, the non-integrable case shows faster complexity growth for the quantum circuit with slower entanglement growth, as seen from Fig.\,\ref{fig:panel_results_1}\,(f), demonstrating a diminished but still exponential advantage.
Importantly, we observe that Fig.\,\ref{fig:MU_variation}\,(b) shows strong dependence of the reachable time on the number of layers $M_U$, which is in agreement with the slope variation shown in Fig.\,\ref{fig:panel_results_1}\,(e).

Lastly, we address the scaling of the total number of gradient steps (Eq.\,\eqref{eq:der_T}) required to time evolve the wave function to the maximum reachable time $t^*$. In Appendix~\ref{appendix:classical_optimization} we show that it scales near-linearly with $N_q$ and with $t^*$, respectively. This means that the full algorithm has a total resource cost that scales polynomially with the time that can be accessed accurately.
}

\subsection{Optimization with a layered environment \label{sec:result_2}}

In the previous section, we have shown that the state-unitary $U$ can be approximated by the layered quantum circuit using exponentially fewer parameters than the {\denseAnsatz} ansatz.
In a real simulation, however, the environments should also be approximated. We now investigate if the environments $| l\rangle, |r\rangle$ can also be represented by layered quantum circuits using fewer parameters.
To address the question, we take the exact environments obtained during the $M_U = 1$ simulation shown previously in Fig.\,\ref{fig:panel_results_1}\,(b) and approximate them with the $M_E$--layer sequential circuits. The approximation is based on alternative update with polar decomposition outlined in Appendix\,\ref{appendix:classical_optimization}.
{ For $g = h = J$,} we plot the approximation error in Fig.\,\ref{fig:max_E_err}\,(a).

We observe that, for a fixed $M_U = 1$, the  environment approximation error increases upon increasing $N_q$, while the error decreases with increasing $M_E$.
We set a threshold $10^{-4}$ in the approximation errors for the environment, which is motivated by the respective error threshold in the fidelity density.
From Fig.\,\ref{fig:max_E_err}\,(a), we see that the error in the environment approximation remains strictly below $10^{-4}$ during the whole time evolution, if it is approximated using $M_E = N_q$ layers. 
It remains an open question on whether the approximation holds for larger $N_q$.

When our observation holds, this allows one to determine the number of variational parameters to approximate the environment.
Since $M_E \propto N_q$, $N_q \propto t^*J$ (see Fig.\,\ref{fig:panel_results_1}\,(b)), and each layer of the sequential circuit for the environment contains $2 N_q - 2$ two-qubit gates, representing the environment requires $\mathcal{O}(t^2)$ or $\mathcal{O}(N_q^2)$ two-qubit gates.
By contrast, representing the environment exactly using dense unitary requires $\mathcal{O}(e^{2 N_q})$, or, equivalently, $\mathcal{O}(e^{2 t J})$ parameters.
The fact that we can approximate the environment with quantum circuits efficiently makes the overall algorithm scaling polynomially in time $t J$ instead of exponentially.
We note that although {\it a priori} the complexity of the environment approximation for the {\ourAnsatz} ansatz is not known, there exist exact solutions representing the environments for infinite brickwall circuits~\cite{gopalakrishnan2019unitary,jobst2022finite}. These exact solutions are formed by contracting $O(N_{\textrm{depth}}^2)$ number of gates, where $N_{\textrm{depth}}$ is the depth of the brickwall circuit, which is consistent with our finding here.

Using the condition $M_E = N_q$, we simulate classically the time evolution algorithm using {\ourAnsatz} with $M_U = 1$, different $N_q$ and layered sequential circuits for environment.
We plot the $\langle \sigma^z \rangle$ expectation value, the reachable time $(tJ)^*$ as a function of $N_q$, and the error in the fidelity density $1 - \mathcal{F}$ in Fig.\,\ref{fig:max_E_err}\,(b).
The reachable time is again determined by the threshold value $\mathcal{F} = 1 - 10^{-4}$.
As shown in the left inset, we observe slightly smaller reachable times, compared to the simulation with the exact environment, due to the accumulation of the approximation errors and approximated environment.
Nevertheless, the reachable time $(tJ)^*(N_q)$ retains the linear scaling with $N_q$.

\subsection{Simulation on QPU}
\label{sec:result_3}

\begin{figure}[t!]
    \centering
    \includegraphics[width=\columnwidth]{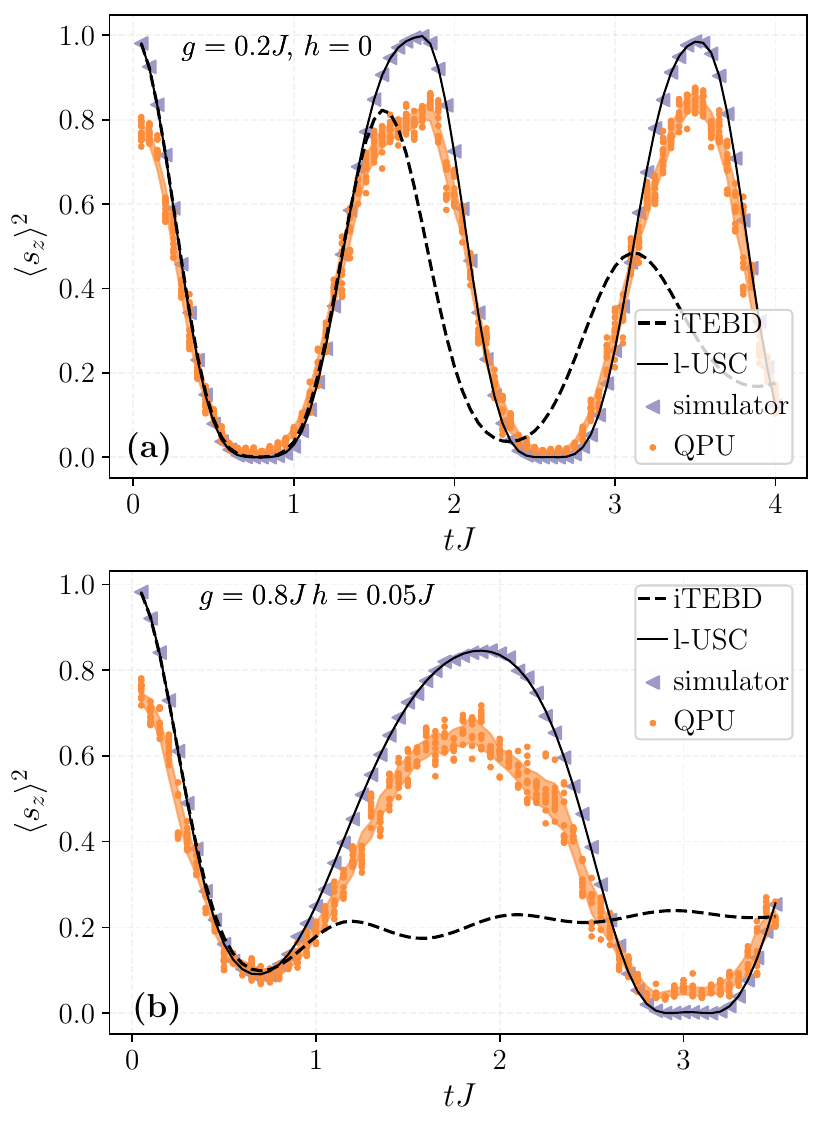}
    \caption{{\bf (a)}: Expectation value of $|\langle \hat{\sigma}^z(t)\rangle|^2$ obtained by various methods: quasi-exact (iTEBD), the presented algorithm with $N_q = 2, M_U = M_E = 1$ ({\ourAnsatz}), simulation of an ideal quantum device (simulator), and on the real hardware \texttt{ibmq-jakarta} (QPU) at $h / J = 0$, $g / J = 0.2$.
    {\bf (b)}: $h / J = 0.05,$ $g / J = 0.8$.}
    \label{fig:QPU}
\end{figure}

Given the available cloud-based QPU from IBM-Q, we implement the circuit shown in Fig.\,\ref{fig:ansatz}\,(d) for $N_q = 2$ and measure $|\langle \sigma^z(t) \rangle|^2$ as the probability of projecting onto the $|00\ldots\rangle$ state.
The parameters of the environments and states are optimized on a classical computer.
Unfortunately, the available hardware does not allow to use controlled two-qubit gates in large amount, since they require decomposition into several non-controlled two-qubit gates.
Due to error and noise levels, this is out of reach for the available device.
This prevents us from measuring the algebraic value of $\langle \sigma^z(t J)\rangle$ using the Hadamard test.
Nevertheless, the numbers of qubits and gates required to run the algorithm until the time $t^*$ scale linearly and quadratically, respectively with $t^*$.
Depending on the device connectivity, an additional constant to linear factor in $t^*$ overhead may occur in implementing the controlled-unitary operation.
Therefore, with improved read-out and gate noise level, we expect this algorithm to be usable on the NISQ devices.
For the measurement of the squared magnitude, we consider two parameter sets: $g = 0.2\,J$, $h = 0$ and $g = 0.8\,J$, $h = 0.05\,J$ and $N_q = 2$, $M_U = M_E = 1$. 
To mitigate the device noise, we employ the {\it randomized circuits averaging} introduced in Ref.\,\cite{Lin_2021} (for details, see Appendix\,\ref{appendix:randomized}), and readout error mitigation.

In Fig.\,\ref{fig:QPU}, we include data obtained from various sources. This includes the quasi-exact simulation (iTEBD), the classical simulation of the algorithm (\ourAnsatz), simulation of the magnetization measurement on a fault-tolerant device using finite number of circuit `shots` (simulator), and, finally, the direct measurement on real hardware device \texttt{ibmq-jakarta} (QPU). 
Due to small expressive power of the quantum circuit at $N_q = 2$, $M_U = M_E = 1$, the exact and the simulated time evolution algorithm results agree only up to $tJ = 1.4$ in the former and $0.8$ in the latter cases. However, the quantum hardware measurement shows a good degree of agreement with the classical simulation of the {\ourAnsatz} ansatz and the simulation of QPU on a classical computer.

\section{Discussion}
\label{sec:discussion}

In this work, we introduced and studied a hybrid quantum-classical algorithm for time evolution { of translationally invariant infinite systems} based on the {\ourAnsatz} ansatz, which is a generalization of the sequential quantum circuit motivated by uniform MPS.
{ We proposed a novel framework for computing the overlap density and expectation values of local observables in the thermodynamic limit based on the new way of constructing the transfer matrix operator.}
Unlike previous works~\cite{Barratt_2021,dborin2022simulating,gopalakrishnan2019unitary}, we construct the transfer matrix in the mixed representation.
In this formalism, the environments are pure states instead of density matrices.
We represent the environments by quantum circuits and determine the variational parameters of these circuits using gradient descent.
Based on the result from classical simulation, we observe that the number of parameters required to accurately represent the state at a given time $t$ scales linearly with $t$, which gives an exponential advantage { compared} to classical algorithms based on MPS. While such scaling is anticipated based on the theoretical prediction~\cite{childs2019nearly}, more interestingly, we observe numerically that the number of parameters required to represent the environment scales quadratically in the evolution time.
This suggest{s} that while the ansatz has a linearly scaling number of parameters with the evolution time, the overall algorithm for simulating time evolution of infinite system has complexity scaling quadratically in the evolution time using quantum computers with a finite number qubits. Importantly, by working directly in the thermodynamic limit, complexity does not scale with the system size $L$ which is in contrast with the $\mathcal{O}(L t)$ complexity scaling required for a finite system simulation~\cite{Lin_2021}.
We emphasize that, unlike Ref.\,\cite{Barratt_2021,Lin_2021}, we consider multi-layered decomposition of the state-unitary with $M_U \geqslant 1$. { As we have seen from Fig.\,\ref{fig:panel_results_1}\,(b, e), Fig.\,\ref{fig:MU_variation}, and Appendix~\ref{appendix:MU_effect} and Appendix~\ref{appendix:sufficiency_proof}, considering $M_U > 1$ leads to improvement in the ansatz performance only for time evolution in the non-integrable $h / J \neq 0$ case.}

We note that we can also perform imaginary time evolution with the proposed algorithm with the price of one additional ancilla qubit~\cite{https://doi.org/10.48550/arxiv.2111.12471} to realize the non-unitary gates in the transfer matrix. One straightforward application is the study of ground states for the infinite systems.
However, important questions remain on whether one would observe similar polynomial advantages~\cite{Haghshenas_2022} in representing the ground state using {\ourAnsatz} for an infinite system, and on whether the environments of the ground states can be efficiently represented as quantum circuits.
Another future direction is to consider the generalization for quantum systems and circuits in two dimensions.
For instance, recently the formal generalization of sequential quantum circuit to finite 2D systems is proposed~\cite{wei2022sequential}. The study of ground states of finite 2D systems are performed using quantum circuits of isometric tensor network states~\cite{slattery2021quantum}.
We believe it will be therefore beneficial to generalize the {\ourAnsatz} ansatz and extending the algorithm to infinite two-dimensional systems.

\begin{acknowledgements}
\noindent We thank Andrew Green for previous works on related topics and discussions.
S.\,L. thanks Ra\'ul Morral Yepesez for helpful discussions.
We acknowledge the use of IBM Quantum services for this work. The views expressed are those of the authors, and do not reflect the official policy or position of IBM or the IBM Quantum team.
Numerical simulations used the high-performance package \texttt{lattice{\_}symmetries}~\cite{westerhout2021latticesymmetries} for quantum state vectors manipulation, and the \texttt{pytorch}~\cite{https://doi.org/10.48550/arxiv.1912.01703} package for the Arnoldi algorithm on a GPU. N.\,A is funded by the Swiss National Science Foundation, grant number: PP00P2{\_}176877. A.\,S. acknowledges support from a research fellowship from the The Royal Commission for the Exhibition of 1851. 
F.\,P. acknowledges the support of the Deutsche Forschungsgemeinschaft (DFG, German Research Foundation) under Germany's Excellence Strategy EXC-2111-390814868. 
The research is part of the Munich Quantum Valley, which is supported by the Bavarian state government with funds from the Hightech Agenda Bayern Plus.

\textbf{Data and materials availability} – Data analysis and simulation codes are available on Zenodo upon reasonable request~\cite{nikita_astrakhantsev_2022_7105374}.

\end{acknowledgements}

\appendix

\section{Correspondence between uniform MPS and {\denseAnsatz}}
\label{appendix:allexact}

In this appendix, we review the equivalence~\cite{schon2005sequential,schon2007sequential} of an MPS of bond dimension $\chi$ and the {\denseAnsatz} ansatz with $N_q = \log_2 \,\chi + 1$ and (ii) demonstrate this correspondence within numerical simulation.

\begin{figure}[t!]
    \centering
    \includegraphics[width=\columnwidth]{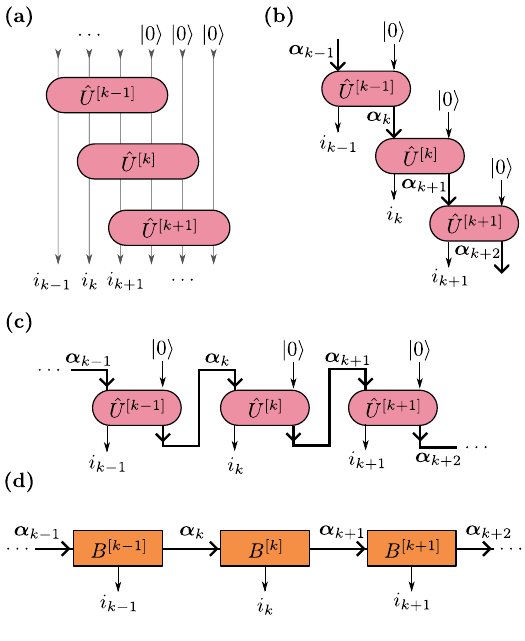}
    \caption{{\bf (a)} A typical {\ourAnsatz} circuit representation of a variational state. {\bf (b)} Same with internal (not physical and not projected) indices denoted as $\boldsymbol \alpha_k$. {\bf (c)} The {\ourAnsatz} circuit rearranged to have a typical horizontal arrangement of uniform MPS. {\bf (d)} The corresponding tensor network representation with uniform MPS $B^{[k]}$. }
    \label{fig:allexact}
\end{figure}


In Fig.\,\ref{fig:allexact}\,(a) we depict a {\denseAnsatz} circuit representing a variational state for $N_q = 4$. Note that each unitary gate $\hat{U}^{[k]}$ has one incoming index contracted to $|0 \rangle$ and one outgoing index being a physical index. The remaining $N_q - 1$ incoming and outgoing indices form composite indices $\boldsymbol \alpha_k$, shown in Fig.\,\ref{fig:allexact}\,(b), where we redraw the {\ourAnsatz} circuit by introducing composite indices $\boldsymbol \alpha_k$.

To see that such circuit is equivalent to an MPS representation with $\chi = 2^{N_q - 1}$, in Fig.\,\ref{fig:allexact}\,(c) we equivalently rearrange the unitaries. To continue further, we note that after contraction, unitaries $\hat{U}^{[k]}$ have the shape $(2, 2^{N_q- 1}, 2^{N_q- 1})$, where the first dimension corresponds to the physical index $i_k$ and the last two correspond to composite indices $\boldsymbol \alpha_k$. Treating them as hidden bonds of an MPS, we recover a tensor-product state matrix $\hat{B}^{[k] i_k}_{a_k b_k}$ with a bond dimension $\chi = 2^{N_q - 1}$, defining a wave function
\begin{gather}
    |\psi\rangle = \sum\limits_{ \{i_k\}} \mbox{Tr}\, \left[ \ldots B^{[k] i_{k}} B^{[k + 1] i_{k + 1}} \ldots\right] |\ldots i_k i_{k + 1} \ldots \rangle,
\end{gather}
where the indices in square brackets $[k]$ enumerate lattice sites, $\alpha_k$ are the virtual bond indices (omitted in $|\psi\rangle$) and $i_k$ are the physical indices enumerating basis states.
This MPS, as we readily observe, is in the right canonical form satisfying 
\begin{gather}
    \label{eq:RCF}
    \sum\limits_{i_k, \alpha_k} B^{[k]i_k}_{\alpha_{k - 1} \alpha_k} \left(B^{[k]i_k}_{\alpha_{k - 1}' \alpha_k} \right)^* = \delta_{\alpha_{k - 1}, \alpha_{k - 1}'}.
\end{gather}

Equivalently, given an MPS with bond dimension $\chi$ in the right canonical form satisfying Eq.\,\eqref{eq:RCF},
we can readily construct a {\denseAnsatz} with unitary  $\hat{U}^{[k]}$ with $N_q = 1 + \log_2 \chi$. To this end, we note that in the right canonical form, $B^{[k]}$ is an isometry mapping from $|\alpha_{k - 1}\rangle$ to $|\alpha_k,\,i_k\rangle$. Any isometry can be rewritten as a unitary acting on a state $|0\rangle$:
\begin{gather}
    B^{[k]} = \hat{U}^{[k]} |0\rangle,
\end{gather}
which represents the contraction of an incoming index of $\hat{U}^{[k]}$ in Fig.\,\ref{fig:allexact}\,(a).
Thus, $\hat{U}^{[k]}$ is the unitary that constructed from $\hat{B}^{[k]}$ and its orthonormal complement.
There is a gauge degree of freedom in chosen the orthonormal complement as only part of the unitary acting on  $|0\rangle$ contribute to the wave function.
We observe equivalent results within iTEBD and our algorithm with {\denseAnsatz} and thus confirm the correctness of the implementation.

\section{Transfer matrix}
\label{appendix:TM}

Transfer matrices are utilized for computation of physical observables and other operations of an infinite system.
In the context of tensor networks, the transfer matrix $\hat{T}$ of the quantum state $| \psi \rangle $ is defined as the repeating block in the computation of the inner product, $\text{Tr}[\hat{T}^N] = \langle \psi | \psi \rangle$, where $N$ is the system size.
Under the mild assumption that the state considered is injective, the expression is well-defined in the thermodynamic limit $N\rightarrow \infty$ regardless of the boundary conditions.
In the following, we extend the same formalism to USC.

\subsection{Transfer matrix of USC}

Utilizing the left and right representations of the {\ourAnsatz} ansatz, we have the freedom to write the inner product as $ \langle \psi_{\text{R}} (\boldsymbol \theta) | \psi_{\text{L}} (\boldsymbol \theta') \rangle = \lim_{N \rightarrow \infty} \text{Tr}[\hat{T}^N]$, of which the repeated block is defined as the transfer matrix shown as the shaded area in Fig.\,\ref{fig:ansatz}\,(a).
Such mixed representation allows us to express the transfer matrix as a linear operator acting on pure states instead of density matrices.

In general, the states $| \psi_{\text{L}} (\boldsymbol \theta') \rangle$ and $| \psi_{\text{R}}(\boldsymbol \theta) \rangle$ can be of different $N_q$ and $M_U$, and they represent similar but not exactly identical states.
We can define the mixed transfer matrix $\hat{T}$ between two different quantum states $|\psi_{\text{L}} \rangle $ and $|\phi_{\text{R}} \rangle $ as the repeating block in the computation of their inner product, $\xi = \langle \phi_{\text{R}} | \psi_{\text{L}} \rangle =  \lim_{N \rightarrow \infty} \text{Tr}\,[\hat{T}^N] $.
The left and right environments $|l \rangle$ and $|r \rangle$ and the leading eigenvalue $\lambda$ are defined similarly to the case of the (not mixed) transfer matrix.
In the thermodynamic limit, the absolute value of the inner product $\lvert \xi \rvert$ is given by
\begin{equation}
    \lim_{N \rightarrow \infty} \lvert \text{Tr}[\hat{T}^N] \rvert = \lim_{N \rightarrow \infty} \lvert \lambda^N \rvert= \begin{cases}
1  \qquad \lvert \lambda \rvert = 1\\
0  \qquad \lvert \lambda \rvert < 1
\end{cases} ,
\end{equation}
which is either an identity or zero, depending on whether the states are identical.
Therefore, a better quantity to consider is instead the overlap density, which is equal to the absolute value of the leading eigenvalue of the transfer matrix $\lvert \lambda \rvert$, satisfying the relation 
\begin{equation}
    \log \lvert \lambda \rvert = \lim_{N \rightarrow \infty} \frac{1}{N}\log{\lvert \xi \rvert}.
\end{equation}

\subsection{Evaluating local observables}

One can evaluate the expectation value of an local observables with respect to the state $|\psi_{\text{L}} (\boldsymbol \theta') \rangle$ following the equation
\begin{equation}
  \langle \hat{\mathcal{O}}\rangle = \frac{\langle \psi | \hat{\mathcal{O}} | \psi \rangle}{\langle \psi |  \psi \rangle}  = \frac{\langle \psi_{\text{L}}(\boldsymbol \theta') | \hat{\mathcal{O}} | \psi_{\text{L}}(\boldsymbol \theta') \rangle}{\langle \psi_{\text{L}}(\boldsymbol \theta') | \psi_{\text{L}}(\boldsymbol \theta') \rangle} .
\end{equation}
The last expression can be evaluated with the right environment in the density matrix form, which satisfies the fixed-point equations.
This is the approach taken by~\cite{Barratt_2021,dborin2022simulating}.
Here, assuming we have the (approximately) identical state in left and right representation, we take an alternative approach and approximate the expression by
\begin{equation}
  \langle \hat{\mathcal{O}}\rangle \approx \frac{\langle \psi_{\text{R}}(\boldsymbol \theta) | \hat{\mathcal{O}} | \psi_{\text{L}}(\boldsymbol \theta') \rangle}{\langle \psi_{\text{R}}(\boldsymbol \theta) | \psi_{\text{L}}(\boldsymbol \theta') \rangle} + O \left(\bigg| \frac{\Delta}{ \lambda_1 - \lambda_2 } \bigg| \right)  .
\end{equation}
The expectation value is now evaluated utilizing the mixed representation of the transfer matrix.
Here, $\Delta$ is the norm difference between the tensors per site describing the states when collapsing the circuit to uniform MPS form, and $\lambda_1$, $\lambda_2$ are the leading and second leading eigenvalues of the transfer matrix of the state $ | \psi_{\text{L}} (\boldsymbol \theta') \rangle$.
Therefore, the expression is exact when $| \psi_{\text{L}} (\boldsymbol \theta') \rangle$ and $| \psi_{\text{R}}(\boldsymbol \theta) \rangle$ are exactly the same state.
The expression is a valid approximation, when the the norm difference is smaller than the size of the gap in the transfer matrix of the physical state  $| \psi_{\text{L}} (\boldsymbol \theta') \rangle$.

In the simpler case where $ | \psi_{\text{L}} (\boldsymbol \theta') \rangle$ and $| \psi_{\text{R}}(\boldsymbol \theta) \rangle$ represent exactly the same physical state, we can always absorb the phase factor into one of the state unitaries such that $\lambda=1$.
The numerator and the denominator are then reduced to
\begin{align}
    \langle \psi_{\text{R}}(\boldsymbol \theta) | \hat{\mathcal{O}} | \psi_{\text{L}}(\boldsymbol \theta') \rangle &= \langle l, 0 | \hat{U}_{\text{R}}^\dagger \hat{\mathcal{O}} \hat{U}_{\text{L}} | 0, r \rangle \\
    \langle \psi_{\text{R}}(\boldsymbol \theta) | \psi_{\text{L}}(\boldsymbol \theta') \rangle &=  \langle l | r \rangle
\end{align}
following the definition of the environments. 
Therefore, we can transform the infinite circuit $\langle \psi_{\text{R}}(\boldsymbol \theta) | \hat{\mathcal{O}} | \psi_{\text{L}}(\boldsymbol \theta') \rangle$ shown in Fig.\,\ref{fig:ansatz}\,(a) into the finite circuit $\langle l, 0 | \hat{U}_{\text{R}}^\dagger \hat{\mathcal{O}} \hat{U}_{\text{L}} | 0, r \rangle$ given in Fig.\,\ref{fig:ansatz}\,(d), which can be implemented on a quantum computer.

If $ | \psi_{\text{L}} (\boldsymbol \theta') \rangle$ and $| \psi_{\text{R}}(\boldsymbol \theta) \rangle$ are not identical, the numerator is suppressed by the additional factor $\lambda^{N-1}$ which cancels out mostly with the additional factor in the denominator $\lambda^{N}$, leading to the expression
\begin{equation}
  \langle \hat{\mathcal{O}}\rangle
  \approx \lim_{N \rightarrow \infty}
  \frac{ \lambda^{N-1} \langle l, 0 | \hat{U}_{\text{R}}^\dagger \hat{\mathcal{O}} \hat{U}_{\text{L}} | r, 0 \rangle}{\lambda^N \langle l | r \rangle}
  = \frac{\langle l, 0 | \hat{U}_{\text{R}}^\dagger \hat{\mathcal{O}} \hat{U}_{\text{L}} | r, 0 \rangle}{\lambda \langle l | r \rangle} .
\end{equation}
The expression suggest that for generic cases, one shall also to take into account the contribution from $\lambda \neq 1$.

\subsection{Derivative of the transfer matrix}
\label{appendix:TM_derivative}

In this appendix, we derive the expression for the derivative of leading eigenvalue of the transfer matrix  with respect to the state unitary.
Consider the transfer matrix $\hat{T}(\boldsymbol \theta)$ depending on variational parameters $\boldsymbol \theta$ and its left and right environments $|l(\boldsymbol \theta)\rangle$, $|r(\boldsymbol \theta)\rangle$, such that $\hat{T}(\boldsymbol \theta) |r(\boldsymbol \theta)\rangle = \lambda(\boldsymbol \theta) |r(\boldsymbol \theta)\rangle$, $\hat{T}^{\dagger}(\boldsymbol \theta) |l(\boldsymbol \theta)\rangle= \lambda^*(\boldsymbol \theta) |l(\boldsymbol \theta)\rangle$.
Therefore, the environments and the leading eigenvalue depend on $\boldsymbol \theta$. We express the leading eigenvalue of the transfer matrix as
\begin{gather}
    \lambda = \frac{\langle l | \hat{T} | r \rangle}{\langle l | r \rangle},
\end{gather}
where the $\boldsymbol \theta$--dependence is omitted for the sake of notation.
Taking derivative with respect to $\boldsymbol \theta$, we obtain
\begin{align}
    \boldsymbol \nabla \lambda &= \frac{\langle \boldsymbol \nabla l| \hat{T} | r\rangle + \langle l| \boldsymbol \nabla \hat{T} | r \rangle + \langle l | \hat{T} | \boldsymbol \nabla r \rangle}{\langle l | r \rangle} \nonumber \\
    &\qquad\qquad - \lambda\frac{\langle \boldsymbol \nabla l | r \rangle + \langle l | \boldsymbol \nabla r \rangle}{\langle l | r \rangle}.
\end{align}

Using $\hat{T} | r \rangle = \lambda |r \rangle$ and $\langle l | \hat{T} = \lambda \langle l|$, we note that the first and third terms in the first fraction cancel out with the second fraction, leaving  
\begin{gather}
    \boldsymbol \nabla \lambda = \frac{\langle l |\boldsymbol \nabla \hat{T} | r \rangle}{\langle l | r \rangle}.
\end{gather}

\subsection{Post-selection probability}
\label{appendix:post_selection_prob}

Consider an arbitrary $N$--qubit vector $|v\rangle$ parameterized by a unitary $\hat{E}_v$ such that $|v\rangle = \hat{E}_v |0\ldots 0\rangle$,
and the (mixed) transfer matrix given by $\hat{T} = \langle 0_{\textrm{last}} |  U_R^\dagger U_L |0_{\textrm{first}}\rangle$, acting on $N$ qubits as shown in Fig.\,\ref{fig:ansatz}\,(c).
The action of the transfer matrix on the vector reads
\begin{align}
    \hat{T} |v\rangle 
    &= \hat{T} \hat{E}_v |0\ldots0_{\textrm{last}} \rangle \nonumber \\
    &=  \langle 0_{\textrm{last}} |  U_R^\dagger U_L  \left(\mathbbm{1}\otimes E_v\right) |0_{\textrm{first}}, 0\ldots0_{\textrm{last}}\rangle.
\end{align}
Therefore, the probability of measuring $| 0_{\textrm{last}} \rangle $, i.\,e., $|0 \rangle$ on the last qubit, is given by
\begin{equation}
    P(0) = \lVert \hat{T} |v\rangle \rVert^2 .
\end{equation}
Note that by definition, the leading eigenvalue of the transfer matrix is unity and the absolute value of the leading eigenvalue of any mixed transfer matrix is equal or less than unity.
%
%

We see that the probability is unity if the input vector is the environment $|r\rangle$, i.\,e., the leading eigenvector  of the transfer matrix $\hat{T}$, since in such case $P(0) = \lVert \hat{T} |r\rangle \rVert^2 = \lvert \lambda \rvert^2 = 1$.
For an arbitrary input state $|v \rangle$, the probability can be expressed as
\begin{equation}
    P(0) = \lVert \hat{T} |v\rangle \rVert^2 = \sum_i \lvert c_i \rvert^2 \lvert \lambda_i \rvert^2,
\end{equation}
where $c_i$ is the coefficient of the eigenbasis of $\hat{T}$.
As a result, the probability $P(0)$ is lower bounded by the square of the coefficient $\lvert c_1 \rvert^2$ corresponding to the leading eigenvector. 
We note that $\lvert c_1 \rvert^2$ is close to unity in the case of Algorithm~\ref{alg:power_method} if we initialize the vector using the environments from the previous time step.

Furthermore, the above property motivates an alternative algorithm for obtaining the environments by maximizing probability $P(0)$ using gradient ascent methods.
This algorithm is potentially more efficient as it only requires the measurement of the last qubit, with gradients measured in absence of ancilla qubits using only the well-known parameter shift rule~\cite{https://doi.org/10.48550/arxiv.1905.13311}.

\section{Details of classical optimization of {\ourAnsatz}}
\label{appendix:classical_optimization}
In this appendix, we provide details on optimization of {\ourAnsatz} that we perform in the course of classical simulation of the time evolution algorithm. 

\subsection{Unitary parametrization and reunitarization}
Optimization of the {\ourAnsatz} ansatz and environments, is performed with the gradient descent method. To incorporate the gradient descent method with the quantum circuits running on a quantum computer, one can employ the decomposition of a general two-qubit gate into 15 gates of the form
\begin{gather}
    \label{eq:singlegate}
\hat{u}(\alpha) = \exp \left(i \alpha \sigma^a \otimes \sigma^b\right),\, a \in \{\hat{I}, \hat{X}, \hat{Y}, \hat{Z}\},
\end{gather}
introduced in Ref.\,\cite{https://doi.org/10.48550/arxiv.1905.13311}.
These gates have an important property, $\partial_{\alpha} \hat{u}(\alpha) = \hat{u}(\alpha + \pi / 2)$, i.\,e. the derivative of these unitary gates is also unitary.
This results in all derivatives of $\hat{U}_{\text{R}/\text{L}}(\boldsymbol \theta)$ or environments being unitary.

In classical optimization, we store $N \times N$ unitaries directly using $2 N^2$ parameters, which is redundant, but significantly speeds-up the optimization.
To ensure correctness of the algorithm, after each finite gradient descent step, we {\it reunitarize} a gate $\hat{U}$ by (i) performing the singular value decomposition $\hat{U} = \hat{V}^{\dagger} \hat{D} \hat{W}$ with $\hat{D}$ being a diagonal matrix with singular values and (2) replacing $\hat{D}$ with a unity matrix: $\hat{U} \to \hat{V}^{\dagger} \hat{W}$.

\subsection{Derivative with respect to a gate}
The main building block of the optimization is obtaining derivatives of the expectation values.
The outlined recipe is applicable not only to two-qubit gates, used in the case of layered state or environment, but also to larger unitaries, used in the optimization of the dense (exact) environment or {\denseAnsatz}.
Consider an expectation value (generally, a scalar complex-valued function) that depends on a set of unitary gates $\lambda(\hat{U}_1, \hat{U}_2, \ldots, \hat{U}_N)$. To compute the derivative with respect to $\hat{U}_k$, any such scalar expectation can be written as $\lambda = \mbox{Tr}\, \hat{U}_k \hat{W}_k^{\dagger}$, with some $\hat{W}_k$ depending on the remaining unitaries. Therefore, the derivative reads:
\begin{gather}
    \label{eq:lambda_derivative}
    \frac{\partial \lambda}{\partial \hat{U}_k} = \hat{W}_k.
\end{gather}

\subsection{Environment optimization}
If the proposed time-evolution algorithm is performed using the a dense representation of environments, the environment is obtained by finding the eigenvector of the transfer matrix  with largest magnitude of eigenvalue using the Arnoldi iteration method.

If the environments are in the layered representation, we first obtain the exact dense environment $|E_{\text{exact}}\rangle$ using the Arnoldi iteration, and then obtain the two-qubit gates of the approximating environment by maximizing the overlap $|\langle E_{\text{layered}}|E_{\text{exact}}\rangle|^2$. 
In such case, instead of performing a gradient descent with the gradient computed using Eq.\,\eqref{eq:lambda_derivative}, we employ the {\it polar decomposition rule}~\cite{Evenbly_2009}.
The polar decomposition rule utilizes the fact that the expectation $\lambda = \mbox{Tr}\, \hat{U}_k \hat{W}_k^{\dagger}$ is maximized over all possible unitaries by taking $\hat{U}_k$ as the reunitarization of $\hat{W}_k$. 
Thus, to optimize a layered environment, we sweep sequentially over all two-qubit gates in $|E_{\text{layered}} \rangle$ and change them using the polar decomposition rule. We stop when the overlap between the environments obtained on two consecutive sweeps exceeds $1 - 10^{-10}$. { For any $N_q$ and at any $t J$, the required number of such sweeps never exceeds 10.}

\begin{figure}[t!]
    \centering
    \includegraphics[width=\columnwidth]{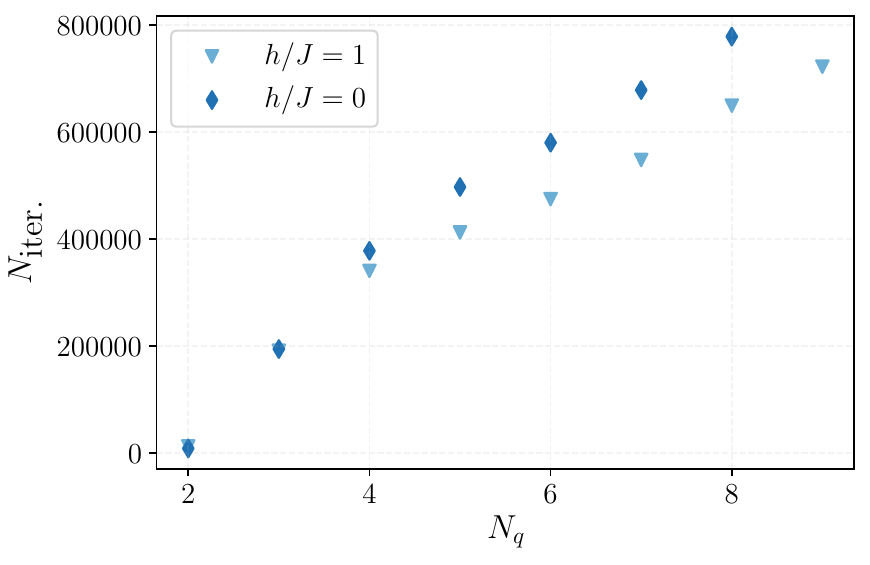}
    \caption{ The total number of the gradient descent iterations $N_{\text{iter.}}(N_q)$ required to reach the time $t^* J$ using the procedure discussed in this this Appendix.}
    \label{fig:n_steps}
\end{figure}

\begin{figure}[b!]
    \centering
    \includegraphics[width=\columnwidth]{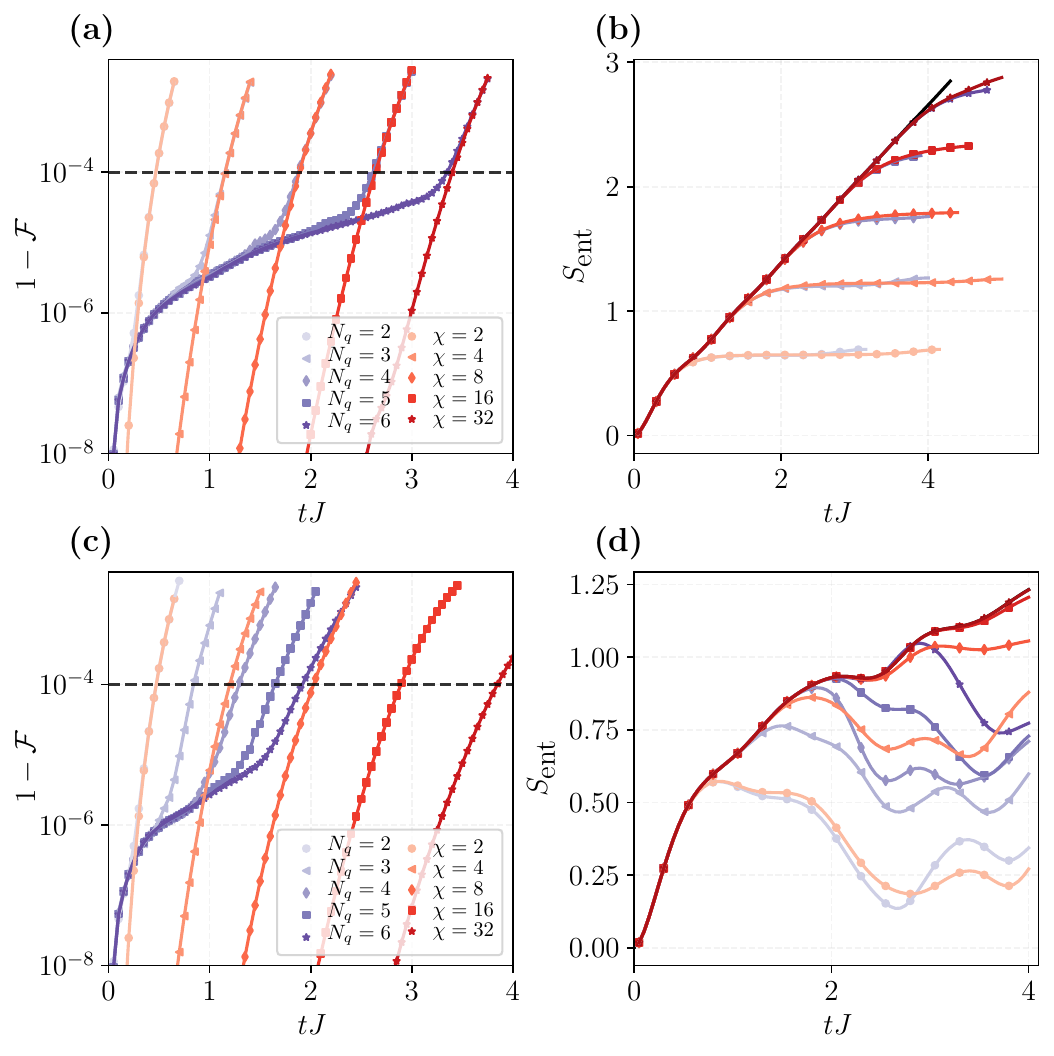}
    \caption{{\bf (a)} Infidelity density between the quasi-exact solution and the {\ourAnsatz} ansatz at $M_U = 1$ or uniform MPS at $\chi = 2^{N_q - 1}$ as a function of $t J$ at $g / J = 1.0$, $h / J = 0.0$. {\bf (b)} Entanglement entropy as a function of $t J$. The solid line represents the quasi-exact solution. { {\bf (c-d)} Same results for the case $g/J = h/J = 1.0$.}} 
    \label{fig:appendix_MUone}
\end{figure}

\subsection{Details of gradient descent method and stopping criteria}
In classical simulation of the proposed algorithm, we employ the redundant parametrization of the unitaries. %
First, having computed the gradient of $|\lambda|^2$ with respect to a unitary $\hat{U}_k$, $\hat{D}_k = \partial |\lambda|^2 / \partial \hat{U}_k = 2 \mbox{Re}\, \left[\lambda^* \partial \lambda / \partial \hat{U}_k \right]$, we project this gradient onto the tangent space of the manifold of $N \times N$ unitary matrices:
\begin{gather}
    \hat{D}_k \to \hat{D}_k - \frac{1}{2} \hat{U}_k \hat{U}_k^{\dagger} \hat{D}_k + \frac{1}{2} \hat{U}_k \hat{D}^{\dagger}_k \hat{U}_k.
\end{gather}
The resulting unitary is reunitarized. 

We employ the ADAM optimizer~\cite{https://doi.org/10.48550/arxiv.1412.6980} with the learning rate $\eta = 3 \times 10^{-3}$.
These two modifications improve the convergence of the algorithm. The optimization finished when the improvement of the leading eigenvalue of the transfer matrix between the two consecutive iterations was less than $10^{-10}$.

Lastly, to speed-up the Arnoldi iteration method, we employed the graphical processing units (GPU) \texttt{Nvidia V100}.

{
\subsection{The total number of gradient iterations}

The number of gradient iterations required to perform time-evolution from time $0$ to $t^*$ with the stopping criteria discussed in this Appendix, is proportional to the total potential hardware run time. Therefore, its scaling is important for the possible future implementation of the outlined algorithm.

In Fig.\,\ref{fig:n_steps}, we show the total number of gradient descent iterations as a function of $N_q$, for the case of the exact environment and non-integrable case $h / J = g / J = 1$. From the data, the scaling is not worse than linear.
}

\section{Sufficiency {(and non-sufficiency)} of the $M_U = 1$ unitary decomposition} 
\label{appendix:MU_effect}

In this appendix, we demonstrate that the accuracy of the time-evolved $M_U = 1$ {\ourAnsatz} ansatz corresponds to the accuracy of the full dense {\denseAnsatz} time-evolved wave function { in the integrable $h / J = 0$ case, which does not hold in the $h / J > 0$ scenario}. More precisely, we show in Appendix~\ref{appendix:sufficiency_proof} that the Gaussian $M_U = 1$ {\ourAnsatz} is exactly equivalent to the Gaussian {\denseAnsatz} for the same $N_q$. To obtain the {\denseAnsatz} ansatz wave function at $N_q$, we optimize the uniform MPS at $\chi = 2^{N_q - 1}$ using the classical iTEBD algorithm. In Fig.\,\ref{fig:appendix_MUone} we show the fidelity densities and entanglement entropies in the both layered and full dense cases.

In Fig.\,\ref{fig:appendix_MUone}\,(a), the infidelity of the uniform MPS differs in the small $tJ$ region, where the infidelity is vanishing, due to different optimization protocols: in the case of the {\ourAnsatz} ansatz, the gradient descent is used, while uniform MPS is optimized using the singular-value decomposition and provides best possible approximation at each time step. 
Nevertheless, the locations of crossing of the $10^{-4}$ infidelity threshold coincide within our resolution. Therefore, increasing $M_U$ would not improve the {\ourAnsatz} ansatz performance, as it is bounded from above by the uniform MPS performance at $\chi = 2^{N_q - 1}.$

Then, in Fig.\,\ref{fig:appendix_MUone}\,(b), we show the entanglement entropy obtained within both the approaches. Similarly, the curves are almost identical in the whole course of time evolution.

\begin{figure}[t!]
    \centering
    \includegraphics[width=\columnwidth]{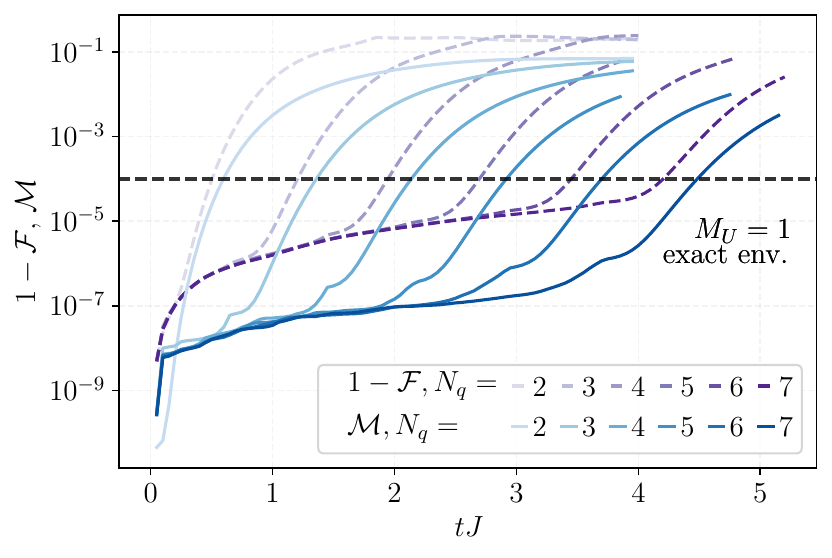}
    \caption{Comparison between the true infidelity density $1 - \mathcal{F}$ computed using the exact solution (crosses) and using the cumulative estimation metric $\mathcal{M}(t) = 1 - \prod_{i < t} |\lambda_{i}|^2$. The simulations were performed at $g / J = 1.0$, $h / J = 0$ with an exact environment and state unitary with $M_U = 1$.}
    \label{fig:results_M_F}
\end{figure}

{
Importantly, as shown in Fig.\,\ref{fig:appendix_MUone}\,(c-d), unlike the $h / J = 0$ case, in the non-integrable $h / J = 1$ setup, $M_U = 1$ is only enough to obtain the full accuracy of the $\chi = 2$ {\denseAnsatz} ansatz, corresponding to the $N_q = 2$ \ourAnsatz. At $N_q > 2$, the {\denseAnsatz} ansatz at the bond dimension $\chi$ shows better accuracy than the {\ourAnsatz} wave function at $N_q = \log_2 \chi - 1$ with $M_U = 1$. This suggests that the integrability is the key element to the sufficiency of $M_U = 1$ in Fig.\,\ref{fig:appendix_MUone}\,(a-b).
}

\section{The accumulated error $\mathcal{M}$ measure}
\label{appendix:M_measure}
During a realistic optimization on a quantum hardware, one has no access to the quasi-exact time-evolved state. Thus, in order to estimate the current error, one can instead define an accumulated error measure
\begin{gather}
    \mathcal{M}(t) = 1 - \prod_{i < t} |\lambda_{i}|^2,
\end{gather}
which is the deviation of the product of leading eigenvalues of the transfer matrices from unity.
Such measure, in case of absence of Trotter errors, should serve as an upper bound for the infidelity $1 - \mathcal{F} \leqslant \mathcal{M}$. 

\begin{figure}[t!]
    \centering
    \includegraphics[width=\columnwidth]{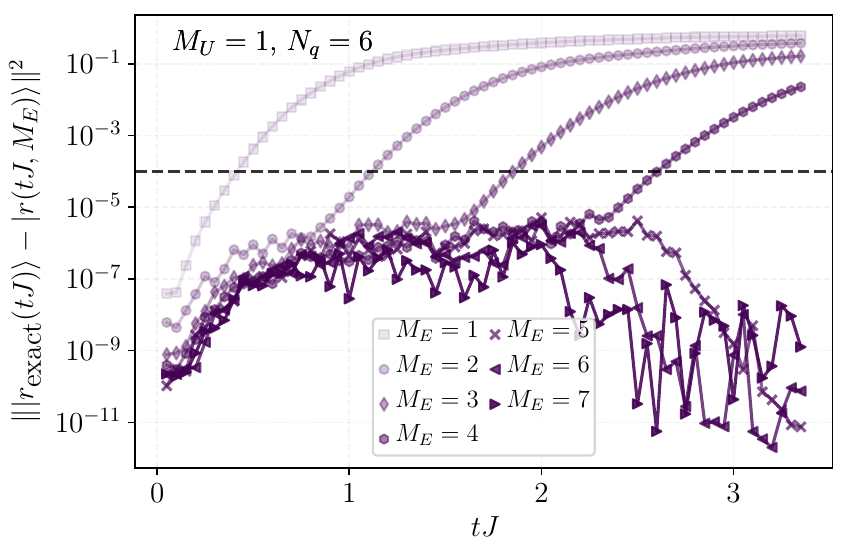}
    \caption{The norm difference between the exact environment $|r_{\text{exact}}\rangle$ and the approximated $|r(M_E)\rangle$ environment at $M_E$ layers as a function of $t J$. The $10^{-4}$ boundary is chosen in the main text as the threshold of {\it accurate representation}. The data is obtained within fitting of the exact environment emerging during the $N_q = 6$ evolution of the {\ourAnsatz} ansatz at $g / J = 1.0$.}
    \label{fig:appendix_represnt}
\end{figure}

However, in this work, we obtain the exact wave function by running a classical iTEBD algorithm at high bond dimension, which breaks translation symmetry to two-site emergent unit cell, while in proposed algorithm we use a second-order translational invariant Trotterization. This discrepancy breaks the inequality, however, the two measures are still strongly correlated. To see this, in Fig.\,\ref{fig:results_M_F} we show the dependence of the true time-evolution infidelity $1 - \mathcal{F}(t)$ and the accumulated error measure. We observe that the two measures cross the $10^{-4}$ threshold at close moments of time.

\begin{figure}[t!]
    \centering
\includegraphics{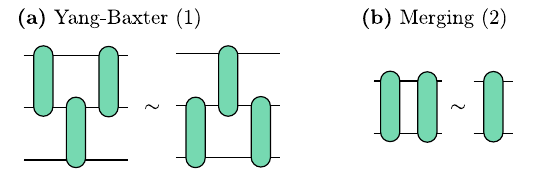}
    \caption{ {\bf (a)} Yang-Baxter equation (1). {\bf (b)} Merging operation (2).}
    \label{fig:YB_merge}
\end{figure}

\begin{figure*}[t!]
    \centering
\includegraphics[width=1.0\textwidth]{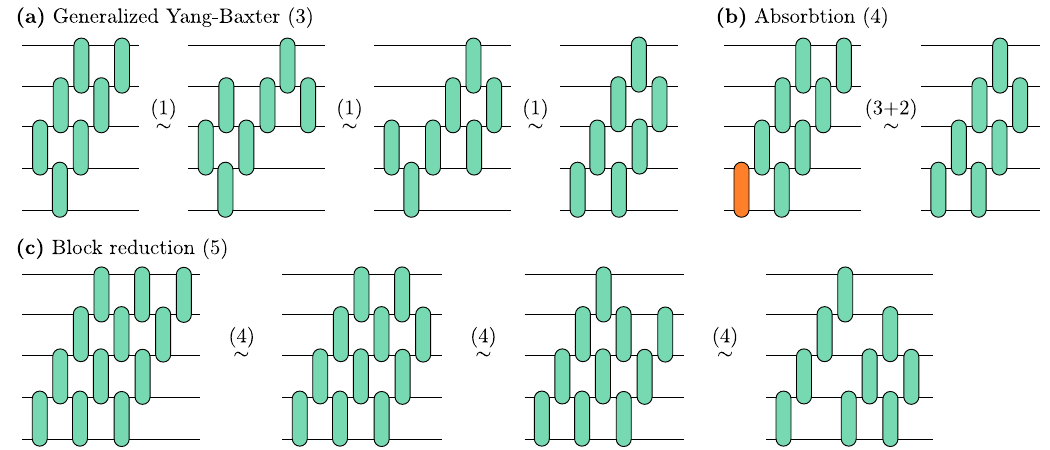}
    \caption{ {\bf (a)} Generalized Yang-Baxter transformation (3). To derive it, we repeatedly apply (1). {\bf (b)} Absorbtion operation (4). After applying (3) to the green gates, the highlighted (orange) gate can be merged using (2). {\bf (c)} Block reduction operation (5) by repeatedly applying (4).}
    \label{fig:other_ops}
\end{figure*}

\section{Environment representation complexity}
In this appendix, we present additional data illustrating the capacity of the layered representation of environment. In Fig.\,\ref{fig:appendix_represnt} we show the norm of discrepancy between the exact environment and the layered environment at $M_E$ layers as a function of $t J$. The exact environments were obtained within the time-evolution of the {\ourAnsatz} ansatz with $N_q = 6$ at $g / J = 1.0$. The maximum evolution time $t J$ is such that the overlap density between the ansatz and the exact state always exceeds $1 - 10^{-4}$.

We observe that, as the complexity of the state grows with time evolution, the approximations $M_E < N_q - 1$ are clearly incapable of accurately representing $|r_{\text{exact}}\rangle$ in the course of time evolution.

\section{Randomized circuits for QPU measurement}
\label{appendix:randomized}
The quantum circuit considered in this paper is described by a series of two-site gates $\{\hat{u}_i\}$. 
When this circuit is implemented on a QPU, the two-qubit gates within \texttt{qiskit} are decomposed into a series of gates
selected from a universal gate set. A small perturbation of a two-site gate may lead to a large change of the decomposition. These differences lead to
large fluctuations in the measured observables due to the QPU noise.

To mitigate these errors, we consider the following procedure. If the two consequent gates $\hat{u}_i$ and $\hat{u}_{i + 1}$ act on the same qubit $q$, we  sample a random $SU(2)$ matrix $\hat{v}$ acting only on the qubit $q$ and modify $\hat{u}_i \to \hat{u}_i \hat{v}$, $\hat{u}_{i + 1} \to \hat{v}^{\dagger} \hat{u}_{i + 1}$. We repeat the measurement scheme in several runs, each time sampling new single-qubit matrices $\hat{v}.$

\section{ Sufficiency of $M_U = 1$ in the integrable case}
\label{appendix:sufficiency_proof}
{

In this Appendix, we explain why in the integrable case $h = 0$, {\ourAnsatz} with $M_U = 1$ is enough to obtain the maximum reachable time available at given $N_q$. 
This includes a proof and a conjecture.
We first show that any Gaussian {\ourAnsatz} with $M_U > 1$ can be reduced to a Gaussian {\ourAnsatz} with $M_U = 1$ of the same $N_q$.
We then discuss the conjecture that in the integrable case $h = 0$, at given $N_q$, the optimal {\denseAnsatz} approximating the time-evolved state is a $N_q$--qubit Gaussian.

A Gaussian {\ourAnsatz} is defined as a {\ourAnsatz} ansatz consisting of two-site Gaussian unitaries, i.\,e., $\hat{U} = e^{i \hat{h}}$ where $\hat{h}$ is a two-site free-fermion Hamiltonian. These unitaries are also known as {\it the matchgates} and we use both terms interchangeably.
Importantly, matchgates are closed under multiplication, i.\,e., remain matchgates, and satisfy the quantum Yang-Baxter equation~\cite{camps2022algebraic}, which is shown in Fig.\,\ref{fig:YB_merge}.
%
%
In the figure, the gates after the $\sim$ sign may have different parameters, but they remain within the matchgates family. We now show that with the merging operation and Yang-Baster equation, we can reduce any $M_U$ circuit to $M_U=1$~\cite{Peng_2022}.

\begin{figure*}[t!]
    \centering
\includegraphics{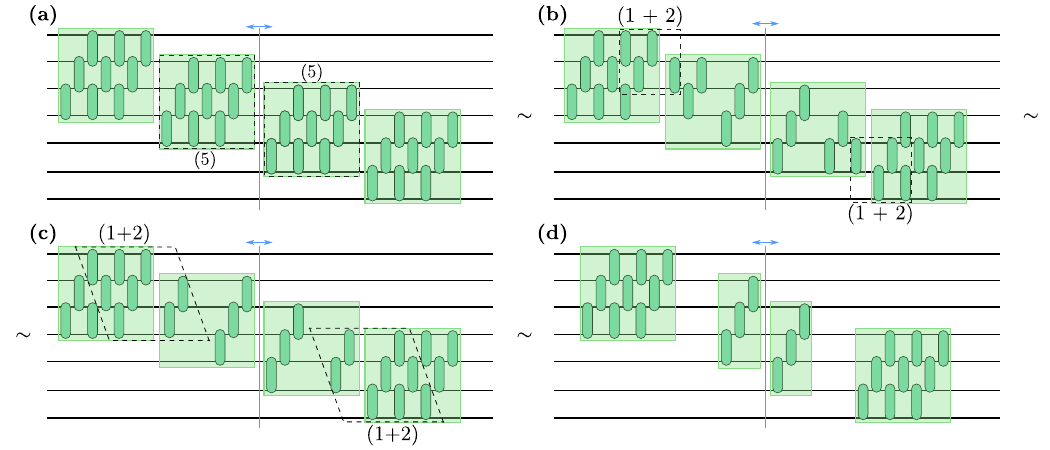}
    \caption{ Graphical proof of the theorem. The blue line represents the pivotal point that indicates with unitaries are being simplified. }
    \label{fig:theorem}
\end{figure*}

To prove that any Gaussian {\ourAnsatz} with $M_U > 1$ can be transformed into $M_U = 1$ case with the same $N_q$, we need additional relations that will be useful. We present these operations in Fig.\,\ref{fig:other_ops}. 
With these operations, we can show by induction that the above statement is true.
As an illustration, in Fig.\,\ref{fig:theorem} we show the $N_q = 4$, $M_U = 3$ ansatz. In the infinite uniform circuit, we select a pivotal point (blue line) and reduce the unitaries adjacent to this point. First, in Fig.\,\ref{fig:theorem}\,(a), we apply the block reduction operation (5) to the unitaries that are reduced. Then, in Fig.\,\ref{fig:theorem}\,(b-c), we apply the Yang-Baxter and Merge operations (1+2) to remove the unitaries beyond a single layer as shown in Fig.\,\ref{fig:theorem}\,(d). Therefore, the unitaries adjacent to the pivotal point can be reduced to the $M_U = 1$ form, and the procedure can be repeated by iteratively moving the pivotal point.

We show that the $M_U=1$ {\ourAnsatz} is equivalent to arbitrary $M_U$ {\ourAnsatz} when the underlying unitaries are matchgates~\cite{Peng_2022,camps2022algebraic}.
The proof further implies that a Gaussian {\denseAnsatz} of a given $N_q$ can be exactly represented as a {\ourAnsatz} with the same $N_q$ and $M_U=1$.
Since a Gaussian unitary gate acting over $N_q$ qubits can be exactly represented as a brickwall circuit using $N_q^3$ matchgates~\cite{Jozsa_2008}, any {\denseAnsatz} composed of Gaussian unitary acting over $N_q$ qubits is equivalent a $M_U=1$ {\ourAnsatz} with the same $N_q$ by folding the circuits using the operations introduced above.

Finally, we conjecture that the optimal {\denseAnsatz} with a fixed $N_q$ approximating the time evolved state $e^{-i\hat{H}t} | 00\ldots 0\rangle$ is Gaussian, where
the integrable Hamiltonian $\hat H$ is defined as in Eq.\,\eqref{eq:hamiltonian} with $h=0$ and a Gaussian {\denseAnsatz} is defined as a {\denseAnsatz} consisting of Gaussian unitaries.
Note that the conjecture implies that the optimal {\ourAnsatz} with a fixed $N_q$ approximating the time evolved state $e^{-i\hat{H}t} | 0\rangle$ is also Gaussian.
This is because that if, at a given $N_q$, there exists a {\ourAnsatz} with some $M_U$ that is non-Gaussian but approximates the state better than the optimal Gaussian {\ourAnsatz}, then there exists a non-Gaussian {\denseAnsatz} of the same $N_q$ that gives a better approximation than Gaussian {\denseAnsatz} of the same $N_q$, which contradict our conjecture.
This conjecture implies that we shall observe the same accuracy for numerical simulation using uniform MPS ({\denseAnsatz}) and the {\ourAnsatz} with $M_U=1$. 
Indeed, for all the numerical simulations performed in this work in the integrable $h = 0$ case as shown in Fig.\,\ref{fig:appendix_MUone} we observed the expected agreement.
}
\begin{widetext}

\section{The equivalence of fixed points}
\label{appendix:identical_fixed_points}

In this appendix, we show that the left and right environments (fixed points), of the {\ourAnsatz} transfer matrix in mixed representation are identical up to complex conjugation.
Because of the formal equivalence between uniform MPS and {\denseAnsatz} shown in Appendix~\ref{appendix:allexact}, we first show such property held in case of uniform MPS.
Consider a uniform MPS in the $\Lambda$--$\Gamma$ canonical form~\cite{Or_s_2008,Schollw_ck_2011}
\begin{equation}
    \label{eq:appendix_FP1}
    \newcommand{\LL}{1}      
\renewcommand{\d}{1.0}   
\renewcommand{\r}{0.3}   
\renewcommand{\a}{0.4}   
|\Psi \rangle = \dots
\begin{diagram}
\draw (0.5,0) -- (1,0);
\draw (1.5,0) circle (0.5);
\draw (1.5,0) node (X) {$\Lambda$};
\draw (2,0) -- (3,0); 
\draw[rounded corners] (3,0.5) rectangle (4,-0.5);
\draw (3.5,0) node {$\Gamma$};
\draw (4,0) -- (5,0);
\draw (5.5,0) circle (0.5);
\draw (5.5,0) node {$\Lambda$};
\draw (6,0) -- (7,0);
\draw[rounded corners] (7,0.5) rectangle (8,-0.5);
\draw (7.5,0) node {$\Gamma$};
\draw (8,0) -- (9,0);
\draw (9.5,0) circle (0.5);
\draw (9.5,0) node {$\Lambda$};
\draw (10,0) -- (11,0);
\draw[rounded corners] (11,0.5) rectangle (12,-0.5);
\draw (11.5,0) node {$\Gamma$};
\draw (12,0) -- (13,0);
\draw (13.5,0) circle (0.5);
\draw (13.5,0) node {$\Lambda$};
\draw (14,0) -- (14.5,0);
\draw (3.5,-.5) -- (3.5,-1);
\draw (7.5,-.5) -- (7.5,-1);
\draw (11.5,-.5) -- (11.5,-1);
\end{diagram}  \dots  ,
\end{equation} 
where $\Lambda$ is a positive-valued diagonal matrix, encoding the Schmidt values.
The combinations of $\Lambda$ and $\Gamma$ give the left (normalized) isometric tensor $A=\Lambda \Gamma$ and the right (normalized) isometric tensor $B=\Gamma \Lambda$.
The overlap between the same physical state and itself is given as the following equation:
\begin{equation}
    \label{eq:appendix_FP2}
    \newcommand{\LL}{1}      
\renewcommand{\d}{1.0}   
\renewcommand{\r}{0.3}   
\renewcommand{\a}{0.4}   
\langle \Psi | \Psi \rangle = \dots
\begin{diagram}
\draw[rounded corners, fill=green!30!teal!30] (2.75,-0.75+\d) rectangle (6.35,0.75+\d);
\draw[rounded corners, fill=green!30!teal!30] (6.75,-0.75+\d) rectangle (10.35,0.75+\d);
\draw[rounded corners, fill=green!30!teal!30] (10.75,-0.75+\d) rectangle (14.35,0.75+\d);
\draw (1.5, 0) node (X) {\phantom{X}};
\draw (0.5,0+\d) -- (1,0+\d);
\draw[fill=white] (1.5,0+\d) circle (0.5);
\draw (1.5,0+\d) node {$\Lambda^*$};
\draw (2,0+\d) -- (3,0+\d); 
\draw[rounded corners, fill=white] (3,0.5+\d) rectangle (4,-0.5+\d);
\draw (3.5,0+\d) node {$\Gamma^*$};
\draw (4,0+\d) -- (5,0+\d);
\draw[fill=white] (5.5,0+\d) circle (0.5);
\draw (5.5,0+\d) node {$\Lambda^*$};
\draw (6,0+\d) -- (7,0+\d);
\draw[rounded corners, fill=white] (7,0.5+\d) rectangle (8,-0.5+\d);
\draw (7.5,0+\d) node {$\Gamma^*$};
\draw (8,0+\d) -- (9,0+\d);
\draw[fill=white] (9.5,0+\d) circle (0.5);
\draw (9.5,0+\d) node {$\Lambda^*$};
\draw (10,0+\d) -- (11,0+\d);
\draw[rounded corners, fill=white] (11,0.5+\d) rectangle (12,-0.5+\d);
\draw (11.5,0+\d) node {$\Gamma^*$};
\draw (12,0+\d) -- (13,0+\d);
\draw[fill=white] (13.5,0+\d) circle (0.5);
\draw (13.5,0+\d) node {$\Lambda^*$};
\draw (14,0+\d) -- (14.5,0+\d);
\draw (3.5,-.5+\d) -- (3.5,-1+\d);
\draw (7.5,-.5+\d) -- (7.5,-1+\d);
\draw (11.5,-.5+\d) -- (11.5,-1+\d);
\draw[rounded corners, fill=red!20] (0.75,-0.75-\d) rectangle (4.35,0.75-\d);
\draw[rounded corners, fill=red!20] (4.75,-0.75-\d) rectangle (8.35,0.75-\d);
\draw[rounded corners, fill=red!20] (8.75,-0.75-\d) rectangle (12.35,0.75-\d);
\draw (0.5,0-\d) -- (1,0-\d);
\draw[fill=white] (1.5,0-\d) circle (0.5);
\draw (1.5,0-\d) node {$\Lambda$};
\draw (2,0-\d) -- (3,0-\d); 
\draw[rounded corners, fill=white] (3,0.5-\d) rectangle (4,-0.5-\d);
\draw (3.5,0-\d) node {$\Gamma$};
\draw (4,0-\d) -- (5,0-\d);
\draw[fill=white] (5.5,0-\d) circle (0.5);
\draw (5.5,0-\d) node {$\Lambda$};
\draw (6,0-\d) -- (7,0-\d);
\draw[rounded corners, fill=white] (7,0.5-\d) rectangle (8,-0.5-\d);
\draw (7.5,0-\d) node {$\Gamma$};
\draw (8,0-\d) -- (9,0-\d);
\draw[fill=white] (9.5,0-\d) circle (0.5);
\draw (9.5,0-\d) node {$\Lambda$};
\draw (10,0-\d) -- (11,0-\d);
\draw[rounded corners, fill=white] (11,0.5-\d) rectangle (12,-0.5-\d);
\draw (11.5,0-\d) node {$\Gamma$};
\draw (12,0-\d) -- (13,0-\d);
\draw[fill=white] (13.5,0-\d) circle (0.5);
\draw (13.5,0-\d) node {$\Lambda$};
\draw (14,0-\d) -- (14.5,0-\d);
\draw (3.5,0.5-\d) -- (3.5,1-\d);
\draw (7.5,.5-\d) -- (7.5,1-\d);
\draw (11.5,.5-\d) -- (11.5,1-\d);
\end{diagram}  \dots   .
\end{equation} 
The transfer matrix in the mixed representation is constructed with the left isometric tensor $A=\Lambda \Gamma$ colored in light red and the right isometric tensor $B^*= \Gamma^* \Lambda^*$ colored in light green.
From the left and right isometric conditions, we see that the left and right environments are simply $\Lambda^*$ and $\Lambda$, respectively.
In this specific case, since the diagonal matrix $\Lambda$ is real and positive, the left and right environments are identical.

In the above case, we consider the isometries with gauge fixing, leading to the $\Lambda$--$\Gamma$ canonical form.
In general, the {\denseAnsatz} is equivalent to uniform MPS in isometric form without gauge fixing.
That is we are allowed to insert identity operators $U^\dagger U=\mathbbm{1}$ and $V^\dagger V = \mathbbm{1}$ to the left and the right of the $\Lambda$ tensor, respectively, where the $U$ and $V$ are arbitrary unitaries.
Similarly, the transfer matrix is constructed by the left and right isometric tensors describing the same physical state, but now, in arbitrary gauge.
The overlap is then given by the equation
\begin{equation}
    \label{eq:appendix_FP3}
    \newcommand{\LL}{1}      
\renewcommand{\d}{1.0}   
\renewcommand{\r}{0.3}   
\renewcommand{\a}{0.4}   
\langle \Psi | \Psi \rangle = \dots
\begin{diagram}
\draw[rounded corners, fill=green!30!teal!30] (3.5,-0.75+\d) rectangle (11.5,0.75+\d);
\draw (1.5, 0) node (X) {\phantom{X}};
\draw (-1.8, 0+\d) -- (-1.6, 0+\d);
\draw[fill=white] (-1.2,0+\d) circle (0.4);
\draw (-1.2,0+\d) node {$U_2^T$};
\draw (-0.8, 0+\d) -- (-0.2, 0+\d);
\draw[fill=white] (0.2,0+\d) circle (0.4);
\draw (0.2,0+\d) node {$U_2^*$};
\draw (0.6,0+\d) -- (1,0+\d);
\draw[fill=white] (1.5,0+\d) circle (0.5);
\draw (1.5,0+\d) node {$\Lambda^*$};
\draw (2,0+\d) -- (2.8,0+\d); 
\draw[fill=white] (2.8,0+\d) circle (0.4);
\draw (2.8,0+\d) node {$V_2^T$};
\draw (3.2, 0+\d) -- (4.2, 0+\d);
\draw[fill=white] (4.2,0+\d) circle (0.4);
\draw (4.2,0+\d) node {$V_2^*$};
\draw (4.6,0+\d) -- (5,0+\d); 
\draw[rounded corners, fill=white] (5,0.5+\d) rectangle (6,-0.5+\d);
\draw (5.5,0+\d) node {$\Gamma^*$};
\draw (6,0+\d) -- (6.4,0+\d);
\draw[fill=white] (6.8,0+\d) circle (0.4);
\draw (6.8,0+\d) node {$U_2^T$};
\draw (7.2, 0+\d) -- (7.8, 0+\d);
\draw[fill=white] (8.2,0+\d) circle (0.4);
\draw (8.2,0+\d) node {$U_2^*$};
\draw (8.6,0+\d) -- (9,0+\d);
\draw[fill=white] (9.5,0+\d) circle (0.5);
\draw (9.5,0+\d) node {$\Lambda^*$};
\draw (10,0+\d) -- (10.8,0+\d); 
\draw[fill=white] (10.8,0+\d) circle (0.4);
\draw (10.8,0+\d) node {$V_2^T$};
\draw (11.2, 0+\d) -- (12.2, 0+\d);
\draw[fill=white] (12.2,0+\d) circle (0.4);
\draw (12.2,0+\d) node {$V_2^*$};
\draw (12.6,0+\d) -- (13,0+\d); 
%
%
\draw[rounded corners, fill=red!20] (-0.5,-0.75-\d) rectangle (7.5,0.75-\d);
\draw (1.5, 0) node (X) {\phantom{X}};
\draw (-1.8, 0-\d) -- (-1.6, 0-\d);
\draw[fill=white] (-1.2,0-\d) circle (0.4);
\draw (-1.2,0-\d) node {$U_1^\dagger$};
\draw (-0.8, 0-\d) -- (-0.2, 0-\d);
\draw[fill=white] (0.2,0-\d) circle (0.4);
\draw (0.2,0-\d) node {$U_1$};
\draw (0.6,0-\d) -- (1,0-\d);
\draw[fill=white] (1.5,0-\d) circle (0.5);
\draw (1.5,0-\d) node {$\Lambda$};
\draw (2,0-\d) -- (2.8,0-\d); 
\draw[fill=white] (2.8,0-\d) circle (0.4);
\draw (2.8,0-\d) node {$V_1^\dagger$};
\draw (3.2, 0-\d) -- (4.2, 0-\d);
\draw[fill=white] (4.2,0-\d) circle (0.4);
\draw (4.2,0-\d) node {$V_1$};
\draw (4.6,0-\d) -- (5,0-\d); 
\draw[rounded corners, fill=white] (5,0.5-\d) rectangle (6,-0.5-\d);
\draw (5.5,0-\d) node {$\Gamma$};
\draw (6,0-\d) -- (6.4,0-\d);
\draw[fill=white] (6.8,0-\d) circle (0.4);
\draw (6.8,0-\d) node {$U_1^\dagger$};
\draw (7.2, 0-\d) -- (7.8, 0-\d);
\draw[fill=white] (8.2,0-\d) circle (0.4);
\draw (8.2,0-\d) node {$U_1$};
\draw (8.6,0-\d) -- (9,0-\d);
\draw[fill=white] (9.5,0-\d) circle (0.5);
\draw (9.5,0-\d) node {$\Lambda$};
\draw (10,0-\d) -- (10.8,0-\d); 
\draw[fill=white] (10.8,0-\d) circle (0.4);
\draw (10.8,0-\d) node {$V_1^\dagger$};
\draw (11.2, 0-\d) -- (12.2, 0-\d);
\draw[fill=white] (12.2,0-\d) circle (0.4);
\draw (12.2,0-\d) node {$V_1$};
\draw (12.6,0-\d) -- (13,0-\d); 
\draw (5.5,0.5-\d) -- (5.5,-0.5+\d);
\end{diagram}  \dots   .
\end{equation} 
As a result, the left isometric tensor is now given by $A=U_1 \Lambda \Gamma U_1^{\dagger}$ colored in light red and the right isometric tensor is given by $B^* = V_2^* \Gamma^* \Lambda^* V_2^T$ colored in light green.
%
%
Similarly, by isometric conditions, the left environment is $U_1^* \Lambda^* V_2^T$ while the the right environment is $U_1 \Lambda V_2^\dagger$.
Therefore, the left environment and the right environment is identical up to complex conjugation.

Since any {\ourAnsatz} state can be viewed as a {\denseAnsatz} state, the statement also applies to the {\ourAnsatz} ansatz: when the {\ourAnsatz} transfer matrix is constructed in the mixed representation, the left and right environments are identical up to complex conjugation.

\end{widetext}

\bibliography{main}
\end{document}